\documentclass[onecolumn,fleqn,usenatbib]{mnras}

\usepackage{newtxtext,newtxmath}

\usepackage[T1]{fontenc}

\DeclareRobustCommand{\VAN}[3]{#2}
\let\VANthebibliography\thebibliography
\def\thebibliography{\DeclareRobustCommand{\VAN}[3]{##3}\VANthebibliography}

\usepackage{soul}
\sethlcolor{red}
\soulregister{\cite}7 
\soulregister{\citep}7 
\soulregister{\citet}7 
\soulregister{\ref}7 
\soulregister{\pageref}7 

\usepackage{graphicx}	
\usepackage{amsmath}	
\usepackage{subcaption}

\title[Long-period planet to HD\,222237]{HD\,222237\,b: a long period super-Jupiter around a nearby star revealed by radial-velocity and Hipparcos-Gaia astrometry\footnote{This paper includes data gathered with the 6.5 meter Magellan Telescopes located at Las Campanas Observatory, Chile.}}

\author[]{Guang-Yao Xiao,$^{1,2}$
Fabo Feng$^{1,2}$,  
Stephen A. Shectman$^{3}$, 
C. G. Tinney$^{4}$,
Johanna K. Teske$^{5}$,
B. D. Carter$^{6}$,
\newauthor H. R. A. Jones$^{7}$,
Robert A. Wittenmyer$^{6}$, 
Mat{\'i}as R. D{\'i}az$^{8}$,
Jeffrey D. Crane$^{3}$, 
Sharon X. Wang$^{9}$,  
J. Bailey$^{4}$,
\newauthor S. J. O'Toole$^{10}$,
Adina~D.~Feinstein$^{11}$\thanks{NHFP Sagan Fellow},
Malena Rice$^{12}$,
Zahra~Essack$^{13}$,
Benjamin~T.~Montet$^{14,15}$,
Avi~Shporer$^{16}$,
\newauthor R. Paul Butler$^{5}$\thanks{E-mail: bluaper@gmail.com}
\\
$^{1}$Tsung-Dao Lee Institute, Shanghai Jiao Tong University, 1 Lisuo Road, Shanghai, 201210, People’s Republic Of China\\
$^{2}$School of Physics and Astronomy, Shanghai Jiao Tong University, 800 Dongchuan Road, Shanghai 200240, People’s Republic of China\\
$^{3}$Observatories of the Carnegie Institution for Science, 813 Santa Barbara Street, Pasadena, CA 91101, USA\\
$^{4}$School of Physics and Australian Centre for Astrobiology, University of New South Wales, Sydney 2052, Australia\\
$^{5}$Earth and Planets Laboratory, Carnegie Institution for Science, 5241 Broad Branch Road, NW, Washington, DC 20015, USA\\
$^{6}$University of Southern Queensland, Centre for Astrophysics, USQ Toowoomba, QLD 4350, Australia\\
$^{7}$Centre for Astrophysics Research, University of Hertfordshire, College Lane, Hatfield, Herts AL10 9AB, UK\\
$^{8}$Las Campanas Observatory, Carnegie Institution of Washington, Colina El Pino, Casilla 601 La Serena, Chile\\
$^{9}$Department of Astronomy, Tsinghua University, Beijing 100084, People’s Republic of China\\
$^{10}$Australian Astronomical Optics, Macquarie University, North Ryde, NSW 1670, Australia\\
$^{11}$Laboratory for Atmospheric and Space Physics, University of Colorado Boulder, UCB 600, Boulder, CO 80309\\
$^{12}$Department of Astronomy, Yale University, New Haven, CT 06511, USA\\
$^{13}$Department of Physics and Astronomy, University of New Mexico, 210 Yale Blvd NE, Albuquerque, NM 87106, USA\\
$^{14}$School of Physics, University of New South Wales, Sydney, NSW 2052, Australia\\
$^{15}$UNSW Data Science Hub, University of New South Wales, Sydney, NSW 2052, Australia\\
$^{16}$Department of Physics and Kavli Institute for Astrophysics and Space Research, Massachusetts Institute of Technology, Cambridge, MA 02139, USA\\
}

\date{Accepted 2024 September 10. Received 2024 August 26; in original form 2023 October 27}

\pubyear{2024}

\begin{document}
\label{firstpage}
\pagerange{\pageref{firstpage}--\pageref{lastpage}}
\maketitle

\begin{abstract}
Giant planets on long period orbits around the nearest stars are among the easiest to directly image. Unfortunately these planets are difficult to fully constrain by indirect methods, e.g., transit and radial velocity (RV).  In this study, we present the discovery of a super-Jupiter, HD\,222237\,b, orbiting a star located $11.445\pm0.002$\,pc away.
By combining RV data, Hipparcos and multi-epoch Gaia astrometry, we estimate the planetary mass to be ${5.19}_{-0.58}^{+0.58}\,M_{\rm Jup}$, with an eccentricity of ${0.56}_{-0.03}^{+0.03}$ and a period of ${40.8}_{-4.5}^{+5.8}$\,yr, making HD\,222237\,b a promising target for imaging using the Mid-Infrared Instrument (MIRI) of JWST. 
A comparative analysis suggests that our method can break the inclination degeneracy and thus differentiate between prograde and retrograde orbits of a companion.
We further find that the inferred contrast ratio between the planet and the host star in the F1550C filter ($15.50\,\mu \rm m$) is approximately $1.9\times10^{-4}$, which is comparable with the measured limit of the MIRI coronagraphs. 
The relatively low metallicity of the host star ($\rm-0.32\,dex$) combined with the unique orbital architecture of this system presents an excellent opportunity to probe the planet-metallicity correlation and the formation scenarios of giant planets.

\end{abstract}




\section{Introduction}
To date more than 5500 exoplanets have been discovered and confirmed via RV, transit, direct imaging, astrometry, and microlensing \citep{Akeson2013}. 
Among them, cold massive Jupiters play a crucial role in shaping the architecture and potential habitability of planetary systems, sparking significant interest in understanding their formation, evolution and dynamics (e.g., \citealt{Stevenson1988,Tsiganis2005}). 
However, finding planets on long period orbits by precision velocity monitoring is painstaking.  At least half an orbital period, preferably more, needs to be observed. For planets with periods of many decades, this can take an entire professional career or longer.  Sufficiently precise velocities solve for all the orbital elements except inclination, so only the minimum mass of a planet ($m_{\rm p}\,{\rm sin}\,i$) can be measured, where $i$ is the unknown inclination angle. 

To extend the temporal baseline and simultaneously solve for the inclination angle, a novel approach that combines RV data with astrometric data from both Hipparcos and Gaia data releases has been developed independently by several groups (e.g., \citealt{Snellen2018,Brandt2018,Feng2019MNRAS,Kervella2019,Xuan2020MNRAS,Van_Zandt2024}). 
For instance, some researchers utilize archival RV data along with proper-motion anomalies between Hipparcos and Gaia astrometry to reveal the 3D stellar reflex motion perturbed by unseen companions and accurately determine the masses of long-period Jupiters (e.g., \citealt{Li2021,Philipot2023,Xiao2023}). However it is important to note that the aforementioned methods generally use a single Gaia data release rather than multiple releases, which may result in unreliable constraints for planets with periods comparable to the time baseline of each satellite. Additionally, using proper-motion anomalies alone can introduce an inclination degeneracy, making it challenging to distinguish prograde ($0\leq i\leq90\degr$) and retrograde ($90\degr<i\leq180\degr$) orbits of a planet \citep{Kervella2020}. The prograde orbit aligns with the direction of increasing position angle on the sky, i.e., anticlockwise direction, while the retrograde orbit corresponds to a clockwise direction (see Figure 2.2 of \citealt{Perryman2018}).   

To overcome the above limitations, we have optimized our analysis to use multiple Gaia data releases by simulating the Gaia epoch data with Gaia Observation Forecast Tool \footnote{\url{https://gaia.esac.esa.int/gost/index.jsp}} (GOST). We then employ a linear astrometric model to fit the synthetic data. By minimising the difference between fitted and catalog astrometry, we are able to uncover the nonlinear reflex motion of a star. This approach has been successfully applied to refine the orbits of cold Jupiters around nearby stars (e.g., $\epsilon\,{\rm Ind\,A\,b}$ and $\epsilon$\,Eridani\,b, \citealt{Feng2023MNRAS}). In this study, we present the discovery of a long-period super-Jupiter, orbiting a metal-poor nearby star HD\,222237 on an eccentric orbit.

HD\,222237 ($=$\,GJ\,902, HIP\,116745) is a K3 dwarf with a $V$ magnitude of 7.09 \citep{Gray2006,Koen2010}, located at a heliocentric distance of $11.445\pm0.002$\,pc \citep{GaiaCollaboration2020}. It has an effective temperature of $T_{\rm eff}=4751\pm139$\,K, a surface gravity of ${\rm log}\,\textsl{g}=4.61\pm0.10$\,dex, a metallicity of $\rm[Fe/H]=-0.32\pm0.02$\,dex, 
a mass of $M_{\star}=0.76\pm0.09\,M_{\sun}$ and a radius of $R_{\star}=0.71\pm0.06\,R_{\sun}$ \citep{Stassun2019}. The chromosphere of the star is slightly active with ${\rm log}R_{\rm HK}^{'}=-4.86$ \citep{Tinney2002}, and Ca \uppercase\expandafter{\romannumeral2} HK emission can be found in its spectra.

This paper is organised as follows. In Section \ref{sec:data_method}, we describe the data and the adopted analysis method. The optimal orbital solution of HD\,222237\,b is presented in Section \ref{sec:results}. The paper concludes with a brief discussion and summary in Section \ref{sec:summary}. 

\section{Data and Methods}
\label{sec:data_method}

\subsection{RV and Astrometry Data}
The precision velocity monitoring of HD\,222237 began in August 1998 with the UCLES echelle spectrometer on the 3.9\,m Anglo-Australian Telescope (AAT; \citealt{Diego1990}). UCLES operated at moderate resolution, R$\sim$45,000. Wavelength calibration was provided by an Iodine cell \citep{Marcy1992}.  The data reduction, including the recovery of the spectrometer point-spread-function, is described in \citet{Butler1996}. Due to its lower resolution (by modern standards), the precision of the AAT/UCLES system was limited to $\sim$ 3 $\rm m\,s^{-1}$.

The High Accuracy Radial-velocity Planet Searcher (HARPS; \citealt{Pepe2000}) mounted on the ESO La Silla 3.6\,m telescope began observing HD\,222237 in 2003. The HARPS spectrograph underwent a major fibre link upgrade at the end of May 2015 \citep{LoCurto2015}.  We distinquish between the ``HARPSpre'' and ``HARPSpost'' data in Fig.~\ref{fig:rv}.  The RVs of HARPS spectra were reduced with the SERVAL pipeline \citep{Zechmeister2018} by \citet{Trifonov2020}, and are publicly available at the HARPS-RVBANK archive \footnote{\url{https://cdsarc.u-strasbg.fr/viz-bin/cat/J/A+A/636/A74}}.

The Carnegie Planet Finder Spectrograph (PFS; \citealt{Crane2006, Crane2008, Crane2010}) mounted on the 6.5\,m Magellan II telescope has been observing HD\,222237 since August 2011, extending the total RV baseline to about 25 yr. As with the AAT/UCLES system, PFS employs an Iodine cell for wavelength calibration to deliver high-precision radial velocities. The upper inflection point of the stellar reflex motion was observed by PFS in 2019 (Fig.~\ref{fig:rv}), which enables a precise characterization of the orbital properties of the planet.
The importance of spectrometer resolution to achieving precise RVs is illustrated by the difference in the quality of the UCLES data relative to HARPS and PFS, which operate at a resolution of 120,000-130,000. All the new RV data used in this work are presented in the Appendix \ref{APPE_A}.

To derive astrometric constraints, we use the Hipparcos epoch data (i.e., intermediate astrometry data, IAD for short) from the new Hipparcos reduction of \citet{vanLeeuwen2007} and Gaia second and third data releases (GDR2 and GDR3; \citealt{GaiaCollaboration2018,GaiaCollaboration2023}), as well as synthetic epoch data from GOST to perform joint analysis with RVs. The Hipparcos IAD and Gaia GOST data mainly comprise the scan angle $\psi$ of the satellite, the along-scan (AL) parallax factor $f^{AL}$, and associated observation epoch at barycenter. Since the Gaia IAD is not available, we use GOST to predict the Gaia observations. The choice of Hipparcos version has negligible impact on our analyses, because we directly model the systematics in Hipparcos IAD using offsets and jitters for a given target (see Appendix \ref{APPE_B}), and we are focusing on the temporal baseline between two satellites ($\rm \sim25\,yr$) when applying for long-period systems \citep{Feng2023MNRAS}.

\subsection{Methods}
The complete methodology of jointly modelling RV and astrometry has been detailed in our previous work \citep{Feng2019MNRAS,Feng2021MNRAS,Feng2023MNRAS}; therefore, we provide a relatively brief introduction about the basic process. Further theoretical formulations can be found in the Appendix \ref{APPE_B}. 

We first model the astrometry of the target system barycenter (TSB) at the GDR3 reference epoch. To solve the problem of perspective acceleration, we transform the above TSB astrometry from the equatorial coordinate system to the Cartesian system to obtain the state vector. The state vector is propagated to the Hipparcos epoch, and we then transform the new vector back to equatorial coordinate system \citep{Lindegren2012,Feng2019ApJS}. Next, we simulate both GDR2 and GDR3 AL abscissae with GOST by adding the stellar reflex motion onto the linear motion of TSB, and fit a five-parameter model to the synthetic abscissae. That fitted astrometry, along with catalog data, is used to construct the likelihood for GDR2 and GDR3. Likewise, we can also model the Hipparcos abscissae and calculate the corresponding likelihood. 
For the RV likelihood, we initially take into account all available noise proxies, e.g., $S$-index of PFS, Bisector inverse span (BIS; \citealt{Queloz2001}) of HARPS and All Sky Automated Survey (ASAS; \citealt{Pojmanski1997}) photometry, and apply a moving average (MA) algorithm to model time-correlated noise in RVs \citep{Feng2017MNRAS}. However, we found this red noise model, compared with white noise model (e.g., jitter term for each instrument), has negligible impact on constraining the orbit in this work. Therefore, we choose the latter to construct the likelihood, which can significantly reduce the free parameters.

With the total likelihood ($\mathcal{L}=\mathcal{L}_{\rm RV}\cdot\mathcal{L}_{\rm Hip}\cdot\mathcal{L}_{\rm Gaia}$), we finally derive the orbital solution by sampling the posterior via the parallel-tempering Markov Chain Monte Carlo (MCMC) sampler \texttt{ptemcee} \citep{Vousden2016}. 
\texttt{ptemcee} is extensively used for sampling from complex, high-dimensional, often multimodal probability distributions. It is capable of traversing different modes at higher temperatures, as well as exploring individual modes at lower temperatures, in order to avoid getting stuck in a local minimum. We employ 30 temperatures, 100 walkers, and 50,000 steps per chain to generate posterior distributions for all the fitting parameters, with the first 25,000 steps being discarded as burn-in.
A Python script that incorporates our complete models (except for the red noise model) is available at \url{https://github.com/gyxiaotdli/mini_Agatha}.

The public package \texttt{orvara} \citep{Brandt2021AJ} was also designed to fit full orbital parameters to any combination of RVs, relative and absolute astrometry. 
It uses the cross-calibrated absolute astrometry from an Hipparcos-Gaia catalog of astrometric accelerations (HGCA; \citealt{Brandt2018}), which corresponds to a single Gaia data release. However, our method with multiple data releases being incorporated is capable of enhancing the orbital constraint due to the inclusion of additional information. Besides, it is important to note that there is uncertainty in the estimation of the calibration parameters between Hipparcos and Gaia \citep{Brandt2018, Lindegren2020}.
Considering this, our method adopts a case-by-case strategy that directly employs jitters and offsets to model astrometric systematics a posteriori. It has been proven effective to avoid the inflation of uncertainties during the frame transformation \citep{Feng2021MNRAS}. Since our method is independent from aforementioned calibration, it can theoretically be applied to extensive Gaia sources whose Hipparcos measurements are not available, particularly for direct imaging systems without accessible RVs (Feng et al, in prep).

To justify the robustness of our detection, we initially conduct a comparative analysis between the widely-used tool \texttt{orvara} (RV+HGCA) and our method without the incorporation of GDR2 (RV+HG3). Then we introduce GDR2 into our model (RV+HG23) and demonstrate the advantage of this inclusion in breaking the inclination degeneracy, thereby differentiating between prograde and retrograde orbits of the planet HD\,222237\,b. 

\section{Results}

\begin{figure}
    \centering    \includegraphics[width=\textwidth]{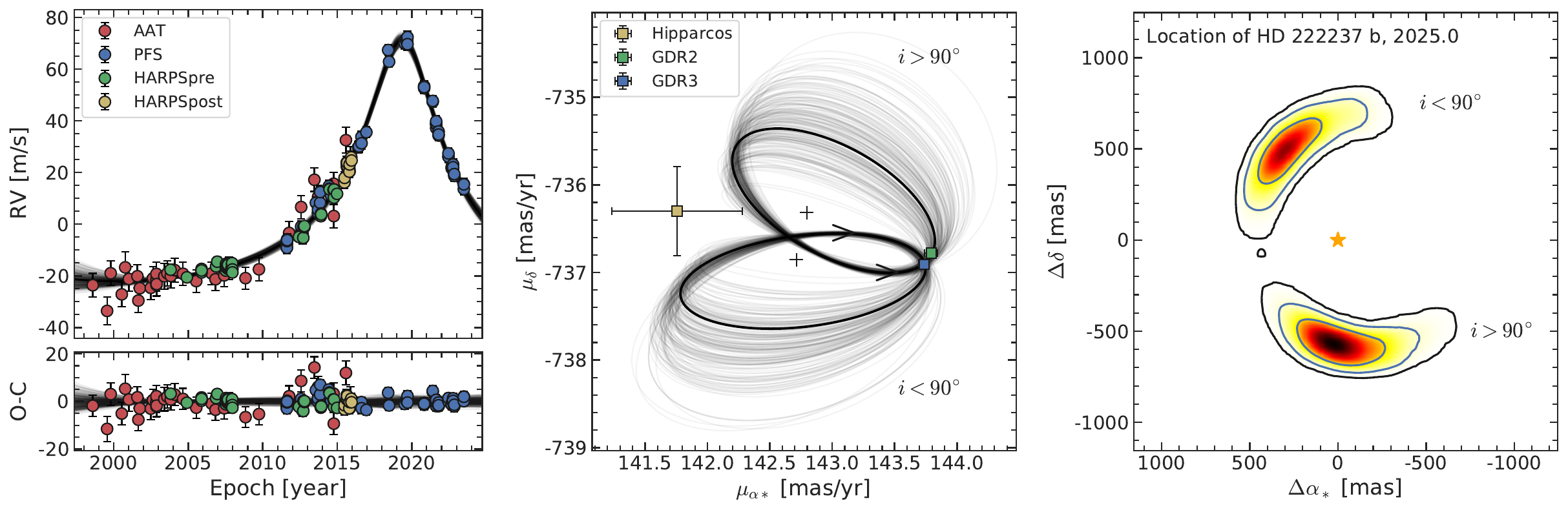}
    \caption{\texttt{orvara} fits to RV and Hipparcos-Gaia astrometry. Left panel: RV curve of HD\,222237\,b. The points with error bar denote the RV measurements and associated uncertainties. The thick black line shows the best-fit orbit. Residuals (O-C) between the observation and the model are plotted below. Middle panel: Astrometric acceleration in right ascension and declination. Two sets of orbits with equivalent likelihood are displayed. The thick lines indicate two best-fit solutions separated by inclination, while the thin line indicate the possible orbital solutions randomly drawn from the MCMC chain. The plus symbols denote the proper motion of TSB, and the arrows indicate the direction in which the proper motion varies over time. The GDR2 astrometry (not used in \texttt{orvara} and RV+HG3 fittings) is added for subsequent analyses. It is evident that two astrometric data points (e.g., Hipparcos and GDR3) can not distinguish two possible solutions, but the inclusion of GDR2 might be helpful to change this situation. Right panel: The predicted position of HD\,222237\,b on January 1st, 2025 and associated $1\,\sigma$, $2\,\sigma$, $3\,\sigma$ uncertainties (contour lines). Two possible sets of contours correspond to prograde ($i<90\degr$) or retrograde ($i>90\degr$) orbits. The orange star denotes the host star HD\,222237. }
    \label{fig:HG3_orvara}
\end{figure}

\label{sec:results}
As shown in Table~\ref{Tab:result}, the primary fitted parameters of our method (both for RV+HG3 and RV+HG23) include the orbital period $P$, RV semi-amplitude $K$, eccentricity $e$, argument of periastron $\omega$ of stellar reflex motion, orbital inclination $i$, longitude of ascending node $\Omega$, mean anomaly $M_{0}$ at the minimum epoch of RV data and five astrometric offsets ($\Delta \alpha*$, $\Delta \delta$, $\Delta \varpi$, $\Delta \mu_{\alpha*}$ and $\Delta \mu_\delta$) of barycenter relative to GDR3. The semi-major axis $a$ of the planet relative to the host, the mass of planet $m_{\rm p}$, and the epoch of periastron passage $T_{\rm p}$ can be derived from above orbital elements. The priors for each parameter are listed in the last column. \texttt{orvara} also adopts \texttt{ptemcee} to fit nine parameters, including the primary star mass $M_{\star}$, the secondary star mass $m_{\rm p}$, $a$, $\sqrt{e}\ {\rm sin}\ \omega$, $\sqrt{e}\ {\rm cos}\ \omega$, $i$, $\Omega$, mean longitude $\lambda_{\rm ref}$ at a reference epoch (2010.0 yr or ${\rm JD}=2455197.50$) and RV jitter (depends on the number of instruments). Some nuisance parameters, such as RV zero point, parallax, and proper motion of system's barycenter, are marginalized by \texttt{orvara} to reduce computational costs. We use the same Gaussian priors for the stellar mass, while use the default priors for the rest (i.e., log-uniform, uniform, geometric, see Table 4 of \citealt{Brandt2021AJ}).

Combining RV and HGCA astrometry (EDR3 version, \citealt{Brandt2021}), \texttt{orvara} yields a planetary mass of ${4.66}_{-0.52}^{+0.63}\,M_{\rm Jup}$, a period of ${37.4}_{-3.8}^{+6.7}$\,yr, an eccentricity of ${0.54}_{-0.03}^{+0.05}$, and two possible inclinations of ${56.5}_{-4.7}^{+5.3}\degr$ and ${123.5}_{-5.3}^{+4.7}\degr$, respectively corresponding to prograde and retrograde orbits. Other fitted and derived parameters are listed in Table~\ref{Tab:result}, while the posterior distributions of selected parameters are displayed in Fig~\ref{fig:corner_orvara} of appendix.
Fig~\ref{fig:HG3_orvara} shows the best-fit Keplerian models to RVs and Hipparcos-Gaia astrometry, and the predicted location of HD\,222237\,b relative to its host star at epochs 2025.0. \texttt{orvara} predicts an angular separation ($\rho$) of $0.59\pm0.05''$ and two possible position angles (PA) of $33\pm32\degr$ and $182\pm16\degr$ in 2025.0. It is evident that \texttt{orvara} can not determine whether the planet is in retrograde or prograde orbital motion.
Similar orbital solutions are also found by RV+HG3 (see Table~\ref{Tab:result}), suggesting the reliability of our method and its consistency with \texttt{orvara}. 
Besides, its posterior distributions for $i$, $\Omega$, $\Delta \alpha*$, $\Delta \delta$, $\Delta \mu_{\alpha*}$ and $\Delta \mu_\delta$ are clearly bimodal (Fig~\ref{fig:corner_HG3}) simply due to the fact that two data points (i.e., Hipparcos and GDR3 absolute astrometry) are insufficient to fully constrain the position and the proper motion of TSB if without a third data point (see the middle panel of Fig~\ref{fig:HG3_orvara}). 

To address above limitations, it is crucial to incorporate GDR2 into our orbital fitting. The optimal orbit of HD\,222237\,b by RV+HG23 gives a slightly longer period of ${40.8}_{-4.5}^{+5.8}$\,yr, an eccentricity of ${0.56}_{-0.03}^{+0.03}$ and a definite inclination of ${49.9}_{-2.8}^{+3.4}\degr$, suggesting a prograde orbital motion. Given the stellar mass of $M_{\star}=0.76\pm0.09\,M_{\odot}$, we derived a mass of ${5.19}_{-0.58}^{+0.58}\,M_{\rm Jup}$ and a semi-major axis of ${10.8}_{-1.0}^{+1.1}$ au for the planet. 
The Root Mean Squares (RMS) of RV residuals for AAT, HARPSpre, HARPSpost and PFS are respectively $\rm 5.20\,m\,s^{-1}$, $\rm 1.72\,m\,s^{-1}$, $\rm 1.97\,m\,s^{-1}$, $\rm 2.13\,m\,s^{-1}$, comparable with the instrument noise. 
We present the posterior distributions of selected orbital parameters in Fig.~\ref{fig:corner_HG23}.

\begin{figure}
    \centering
	\includegraphics[width=0.5\textwidth]{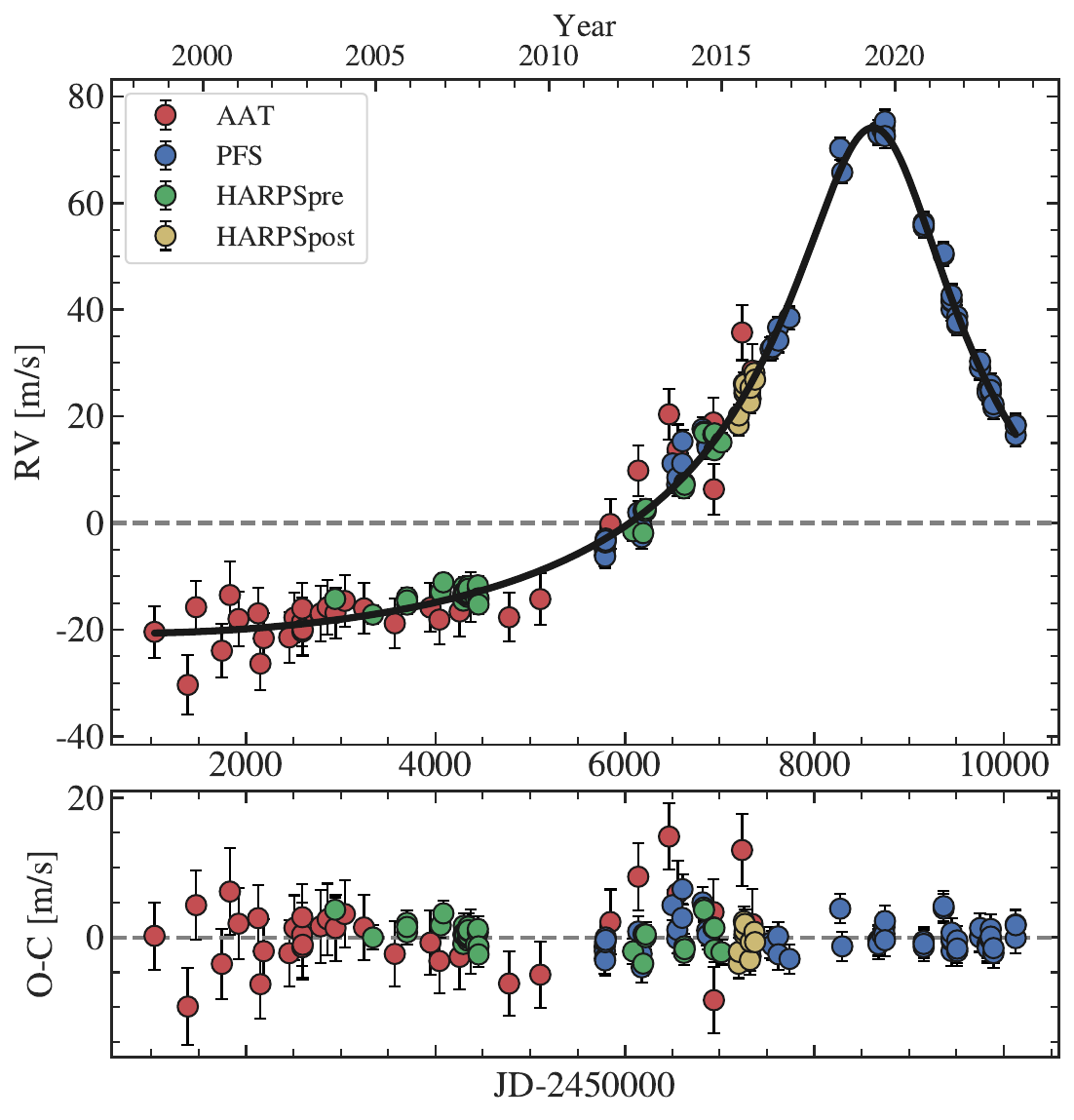}
    \caption{RV+HG23 fits to RVs. The symbols are the same as the left panel of Fig~\ref{fig:HG3_orvara}.}
    \label{fig:rv}
\end{figure}

\begin{figure*}
    \centering
	\includegraphics[width=\textwidth]{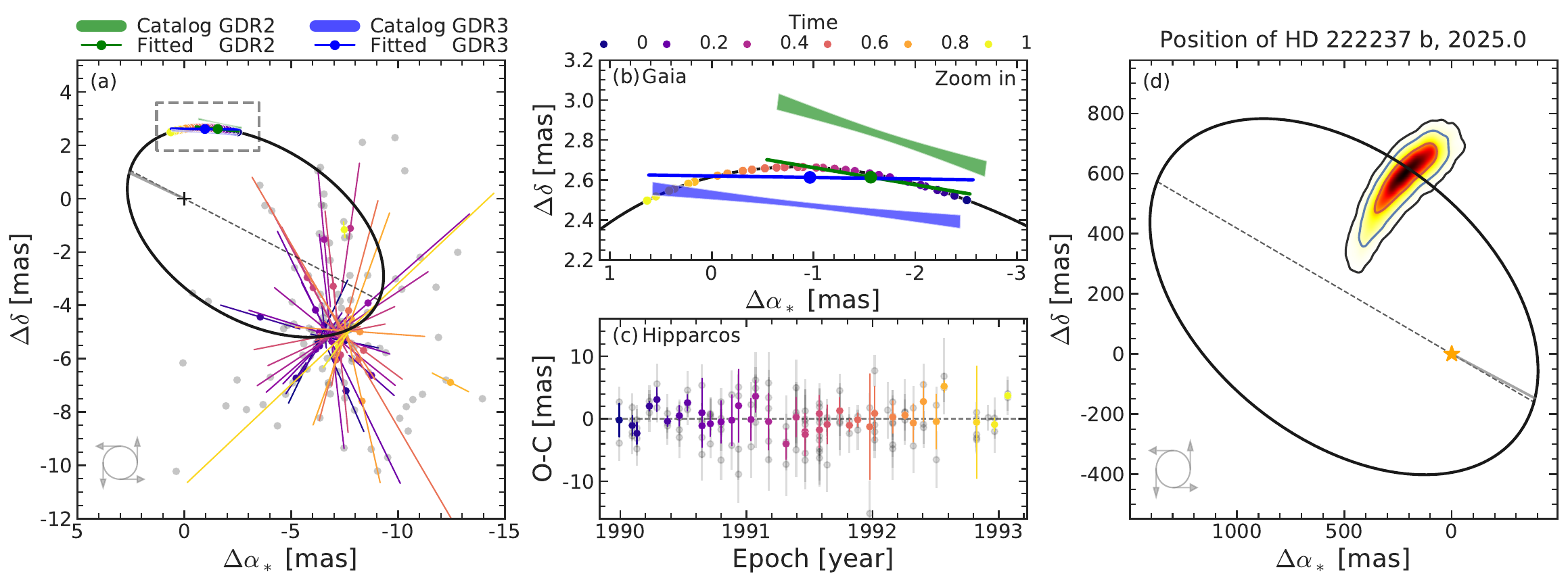}
    \caption{RV+HG23 fits to Hipparcos and Gaia astrometry. (a) The best-fit astrometric orbit of HD\,222237. The black dashed line inside the orbit is the line of nodes joining the ascending node and the descending node. The plus symbol denotes the system's barycenter, and the grey line connects it with the periapsis. The post-fit Hipparcos abscissa residuals are projected into the R.A. and decl. axes (grey dots). Their multiple measurements of each epoch have been binned to single points with colors, and the brightness of colors gradually increases with observation time (the temporal baseline of each satellite is set to 1). The orientations of the error bars of each point denote the along-scan direction of Hipparcos. (b) Zoom in on the rectangle region of panel (a) where depicts the best fit to Gaia GOST data and the comparison between best-fit and catalog astrometry (positions and proper motions) at GDR2 and GDR3 reference epochs. The shade regions represent the uncertainty of catalog positions and proper motions after removing TSB motion. The dot and slope of two lines (blue and green) indicate the best-fit position and proper motion offsets induced by the planet. (c) The residual (O-C) of Hipparcos abscissa. (d) The predicted position of HD\,222237\,b on January 1st, 2025 and associated $1\,\sigma$, $2\,\sigma$, $3\,\sigma$ uncertainties (contour line). The curl at the lower left corner denotes the orientation of the orbital motion.}
    \label{fig:orbit}
\end{figure*}

\begin{figure*}
    \centering
	\includegraphics[width=0.8\textwidth]{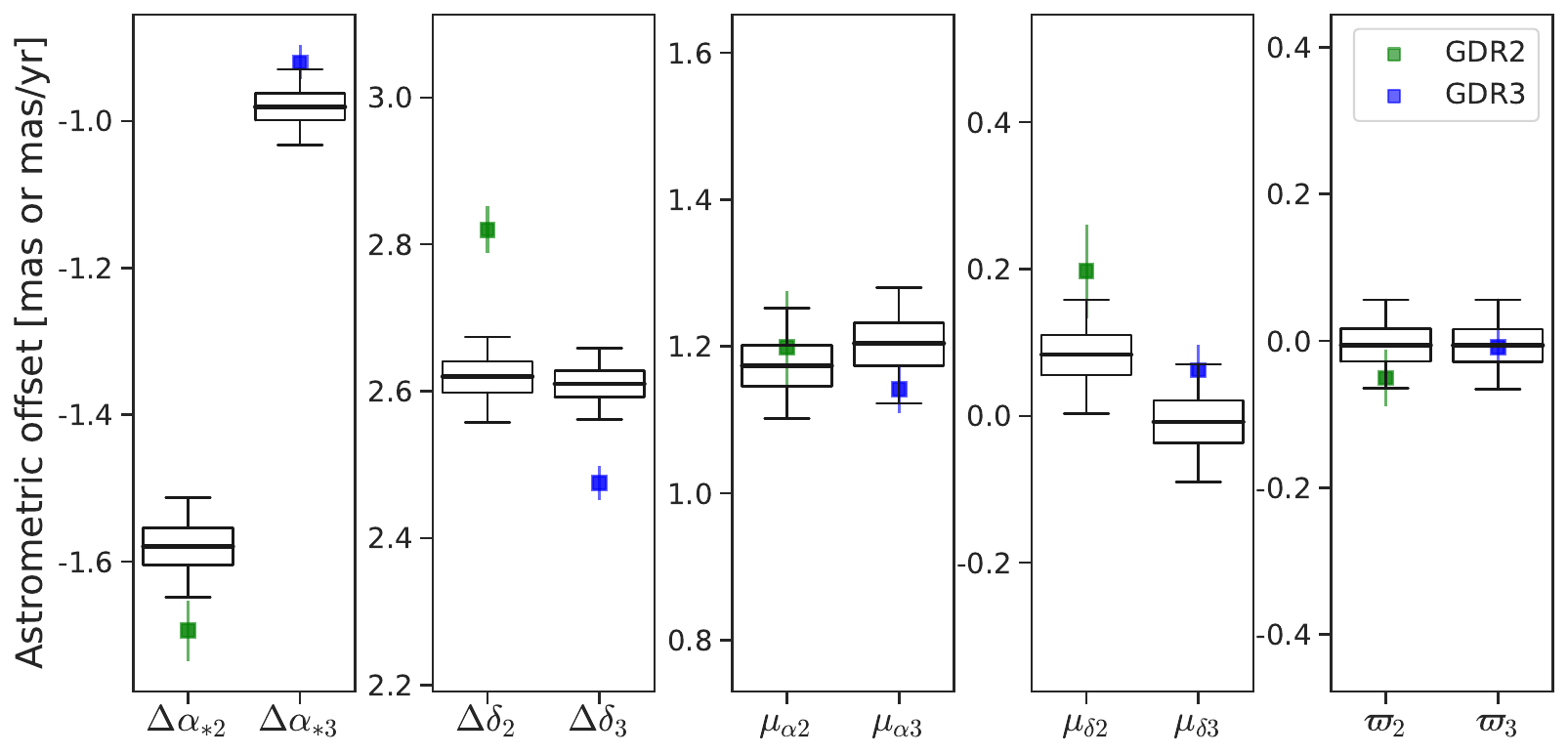}
    \caption{Comparing the five-parameter astrometry of the model prediction to GDR2 and GDR3 astrometry. The barycentric motion has been subtracted for both catalog Gaia data (square) and the prediction (boxplot). The inner thick line, the body and the edge of the boxplot respectively indicate the median, $1\,\sigma$ uncertainty and $3\,\sigma$ uncertainty. 
    The subscripts of the label of the x-axis correspond to the Gaia release number.}
    \label{fig:five_p}
\end{figure*}

Fig.~\ref{fig:rv} and Fig.~\ref{fig:orbit} depict the optimal orbital solution for HD\,222237\,b based on the MCMC posterior of RV+HG23. The former shows the best fit to RVs, while the latter shows the best fit to Hipparcos IAD and Gaia GOST data, and the predicted position of the planet.
In Fig.~\ref{fig:orbit} (a), we project the Hipparcos abscissae along the R.A. and decl. directions for visualization purposes, and encode the observation time with colors. 
In Fig.~\ref{fig:orbit} (b), we use segments and shaded region to visualize the Gaia catalog astromery and the best-fit astrometry. 
All of them have been corrected according to the TSB astrometry. The center of the segment denotes the offset in R.A and decl. relative to the TSB, and the slope denotes the ratio of the proper motion offsets (PMo) in decl. and R.A, and the length is the product of PMo and the temporal baseline of GDR2 or GDR3. 
The fitted GDR2 and GDR3 shown in this panel are determined by fitting a five-parameter model to the synthetic data. Fig.~\ref{fig:orbit} (c) plot the 1D residual of Hipparcos abscissa between the observations and the best fitting. In the last panel of Fig.~\ref{fig:orbit}, we predict the position of the planet on January 1st, 2025. The estimated angular separation is $0.64\pm0.04''$, and the position angle is $21\pm10\degr$, consistent with the prediction based on \texttt{orvara} solution. This planet will reach its maximum angular separation of $1.45\pm0.18''$ in January 2040.
In addition, we present a more intuitive comparison of our predictions with the five-parameter astrometry of GDR2 and GDR3 in Fig.~\ref{fig:five_p}. Overall, the fitting to GDR3 is better than GDR2 due to the longer temporal baseline.

As shown in Table~\ref{Tab:result}, almost all the parameters (prograde orbital solution) from \texttt{orvara} and RV+HG3 are in great agreement within $1\,\sigma$ with the solution obtained by RV+HG23. 
While with the inclusion of GDR2, our method can resolve the TSB ambiguity and is able to differentiate between prograde and retrograde orbits. To further corroborate this conclusion, we inject the posteriors of RV+HG3 into RV+HG23 model and inspect whether the two sets of orbital solution from the former can be distinguished by the latter (Fig~\ref{fig:fourpics}). It can be found that the higher inclination (corresponding to the retrograde orbital solution) will be rejected by RV+HG23 model, suggesting the precision of Gaia, along with the baseline between GDR2 and GDR3 are sufficient to obtain an unambiguous orbital orientation of HD\,222237\,b.
It also should be noted that the use of multiple Gaia DRs does not significantly improve the constraint on long-period orbits, but it can provide additional information about the raw abscissae and thus improve the accuracy of the orbital solutions. 

\begin{table*}
\centering
\caption{Parameters for HD\,222237}\label{Tab:result}
\begin{tabular*}{\textwidth}{@{}@{\extracolsep{\fill}}lcccccc@{}}
\hline \hline
Parameter & Unit & Meaning & RV+HGCA${\rm ^d}$ &RV+HG3&RV+HG23 & Prior${\rm ^e}$ \\
\hline
$P$&day&Orbital period&${13648}_{-1397}^{+2460}$ &${13262}_{-1165}^{+1319}$&${14892}_{-1655}^{+2112}$& Log-$\mathcal{U}(-1,16)$\\
$K$&m s$^{-1}$&RV semi-amplitude&${46.82}_{-0.95}^{+1.3}$ &${46.57}_{-0.82}^{+0.83}$&${47.4}_{-1.0}^{+1.1}$& $\mathcal{U}(10^{-6},10^{6})$\\
$e$&---&Eccentricity&${0.54}_{-0.03}^{+0.05}$ &${0.53}_{-0.03}^{+0.03}$&${0.56}_{-0.03}^{+0.03}$& $\mathcal{U}(0,1)$\\
$\omega$&deg&Argument of periapsis${\rm ^a}$&${3.1}_{-1.3}^{+1.4}$ &${3.2}_{-1.4}^{+1.4}$&${2.6}_{-1.4}^{+1.3}$& $\mathcal{U}(0,2\pi)$\\
$M_0$&deg&Mean anomaly at JD 2451034&${269}_{-11}^{+14}$(J2010) &${153}_{-20}^{+19}$&${176}_{-23}^{+23}$& $\mathcal{U}(0,2\pi)$\\
$i$&deg&Inclination&${56.5}_{-4.7}^{+5.3}$ (${123.5}_{-5.3}^{+4.7}$) &${57.8}_{-3.8}^{+4.9}$ (${122.9}_{-4.5}^{+3.3}$)&${49.9}_{-2.8}^{+3.4}$& Cos$i$-$\mathcal{U}(-1,1)$\\
$\Omega$&deg&Longitude of ascending node&${74.8}_{-8.5}^{+7.9}$ (${137.6}_{-8.3}^{+8.9}$) &${76.0}_{-6.1}^{+7.4}$ (${135.3}_{-6.9}^{+5.6}$)&${69.8}_{-5.7}^{+6.7}$& $\mathcal{U}(0,2\pi)$\\\hline

$\Delta \alpha*$&mas&$\alpha*$ offset&---&${-0.45}_{-0.29}^{+0.26}$ (${-1.38}_{-0.36}^{+0.39}$)&${-0.79}_{-0.36}^{+0.34}$& $\mathcal{U}(10^{-6},10^{6})$\\
$\Delta \delta$&mas&$\delta$ offset&---&${1.87}_{-0.37}^{+0.33}$ (${-1.40}_{-0.14}^{+0.18}$)&${2.43}_{-0.36}^{+0.33}$& $\mathcal{U}(10^{-6},10^{6})$\\
$\Delta \mu_{\alpha*}$&mas\,yr$^{-1}$&$\mu_{\alpha*}$ offset &${1.045}_{-0.090}^{+0.12}$ (${0.918}_{-0.048}^{+0.062}$) &${1.021}_{-0.083}^{+0.081}$ (${0.908}_{-0.044}^{+0.037}$)&${1.13}_{-0.10}^{+0.10}$& $\mathcal{U}(10^{-6},10^{6})$\\
$\Delta \mu_\delta$&mas\,yr$^{-1}$&$\mu_\delta$ offset&${-0.044}_{-0.068}^{+0.087}$ (${-0.53}_{-0.14}^{+0.11}$) &${-0.050}_{-0.064}^{+0.063}$ (${-0.492}_{-0.083}^{+0.095}$)&${0.041}_{-0.068}^{+0.065}$& $\mathcal{U}(10^{-6},10^{6})$\\
$\Delta \varpi$&mas&$\varpi$ offset&$\sim0$&${-0.005}_{-0.018}^{+0.018}$&${-0.009}_{-0.022}^{+0.022}$& $\mathcal{U}(10^{-6},10^{6})$\\\hline
$P$&yr&Orbital period &${37.4}_{-3.8}^{+6.7}$ &${36.3}_{-3.2}^{+3.6}$&${40.8}_{-4.5}^{+5.8}$& ---\\
$a$&au&Semi-major axis${\rm ^b}$ &${10.23}_{-0.83}^{+1.3}$ &${9.99}_{-0.71}^{+0.78}$&${10.8}_{-1.0}^{+1.1}$& ---\\
$m_{\rm p}$&$M_{\rm Jup}$&Companion mass &${4.66}_{-0.52}^{+0.63}$ &${4.56}_{-0.49}^{+0.51}$&${5.19}_{-0.58}^{+0.58}$& ---\\
$T_{\rm p}-2400000$&JD&Periapsis epoch&${45001}_{-2472}^{+1403}$ &${45387}_{-1328}^{+1173}$&${43747}_{-2116}^{+1661}$& ---\\
\hline
$N^{\rm AAT}$&---&AAT observations${\rm ^c}$&35&35&35&---\\
$N^{\rm HARPSpre}$&---&HARPSpre observations &43&43&43&---\\
$N^{\rm HARPSpost}$&---&HARPSpost observations &10&10&10&---\\
$N^{\rm PFS}$&---&PFS observations&63&63&63&---\\
$\gamma^{\rm AAT}$&m\,s$^{-1}$&RV offset for AAT&${13.2}_{-1.9}^{+3.2}$&${13.7}_{-1.9}^{+1.9}$&${11.5}_{-2.8}^{+2.4}$&$\mathcal{U}(10^{-6},10^{6})$\\
$\gamma^{\rm HARPSpre}$&m\,s$^{-1}$&RV offset for HARPSpre&${8.1}_{-2.5}^{+3.7}$&${8.8}_{-2.2}^{+2.1}$&${6.2}_{-3.0}^{+2.8}$&$\mathcal{U}(10^{-6},10^{6})$\\
$\gamma^{\rm HARPSpost}$&m\,s$^{-1}$&RV offset for HARPSpost&${-22.2}_{-2.5}^{+3.7}$&${-21.5}_{-2.3}^{+2.3}$&${-24.1}_{-3.0}^{+2.9}$&$\mathcal{U}(10^{-6},10^{6})$\\
$\gamma^{\rm PFS}$&m\,s$^{-1}$&RV offset for PFS&${-27.9}_{-2.6}^{+3.8}$&${-27.2}_{-2.3}^{+2.2}$&${-29.9}_{-3.0}^{+2.9}$&$\mathcal{U}(10^{-6},10^{6})$\\
$J^{\rm AAT}$&m\,s$^{-1}$&RV jitter for AAT&${4.89}_{-0.67}^{+0.80}$&${5.07}_{-0.73}^{+0.83}$&${4.91}_{-0.69}^{+0.83}$&$\mathcal{U}(0,10^{6})$\\
$J^{\rm HARPSpre}$&m\,s$^{-1}$&RV jitter for HARPSpre&${1.68}_{-0.19}^{+0.23}$&${1.70}_{-0.19}^{+0.23}$&${1.72}_{-0.19}^{+0.23}$&$\mathcal{U}(0,10^{6})$\\
$J^{\rm HARPSpost}$&m\,s$^{-1}$&RV jitter for HARPSpost&${2.13}_{-0.44}^{+0.66}$&${2.28}_{-0.50}^{+0.75}$&${2.27}_{-0.48}^{+0.71}$&$\mathcal{U}(0,10^{6})$\\
$J^{\rm PFS}$&m\,s$^{-1}$&RV jitter for PFS&${1.99}_{-0.21}^{+0.24}$&${2.03}_{-0.22}^{+0.25}$&${2.02}_{-0.22}^{+0.25}$&$\mathcal{U}(0,10^{6})$\\
$J^{\rm hip}$&mas&Jitter for hipparcos&---&${0.75}_{-0.48}^{+0.53}$&${0.80}_{-0.50}^{+0.51}$&$\mathcal{U}(0,10^{6})$\\
$S^{\rm gaia}$&---&Error inflation factor&---&${1.00}_{-0.10}^{+0.10}$&${1.59}_{-0.07}^{+0.07}$&$\mathcal{N}(1,0.1)$\\
\hline 
\multicolumn{7}{l}{$\rm ^a$ The argument of periastron of the stellar reflex motion, differing by $\pi$ with planetary orbit, i.e., $\omega_{\rm p}=\omega+\pi$}.\\
\multicolumn{7}{l}{$\rm ^b$ The semi-major axis $a$ and planet mass $m_{\rm p}$ are derived from fitted parameters assuming the stellar mass as a Gaussian prior.}\\
\multicolumn{7}{l}{$\rm ^c$ These four rows show the number of observations from each spectrometer.}\\
\multicolumn{7}{l}{$\rm ^d$ The orbital solution from \texttt{orvara}. Since some free parameters used by \texttt{orvara} differ from ours, we derive their values using MCMC posteriors.}\\
\multicolumn{7}{l}{$\rm ^e$ The priors of our method, while priors for \texttt{orvara} are listed in \citet{Brandt2021AJ}. Log-$\mathcal{U}(a, b)$ is the logarithmic uniform distribution between $a$ and $b$, }\\
\multicolumn{7}{l}{ Cos$i$-$\mathcal{U}(a, b)$ is the cosine uniform distribution between $a$ and $b$, and $\mathcal{N}(a, b)$ is the Gaussian distribution with mean $a$ and stardare deviation $b$.}\\
\end{tabular*}
\end{table*}

\section{Discussion and Summary}
\label{sec:summary}

\begin{figure*}
    \centering
	\includegraphics[width=0.9\textwidth]{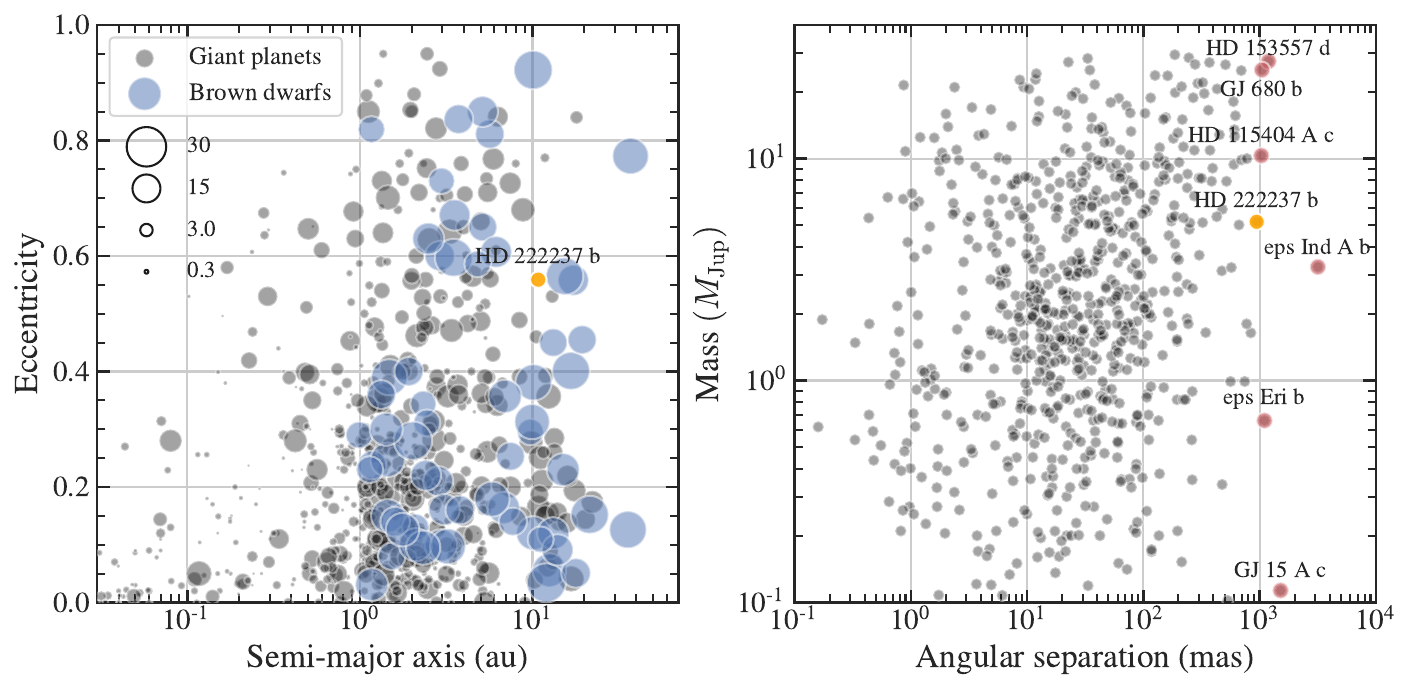}
    \caption{Left panel: Eccentricity versus semi-major axis for RV-detected companions. The size of circles is proportional to the planetary (minimum) mass. Right panel: Planetary (minimum) mass versus angular separation. HD\,222237\,b is marked by an orange circle. We also show six substellar companions with orbits at the widest separation. All the data points are compiled from the NASA exoplanet archive \citep{Akeson2013} on July 1, 2024.}
    \label{fig:mass_ecc_au}
\end{figure*}

\begin{figure}
    \centering
	\includegraphics[width=0.5\textwidth]{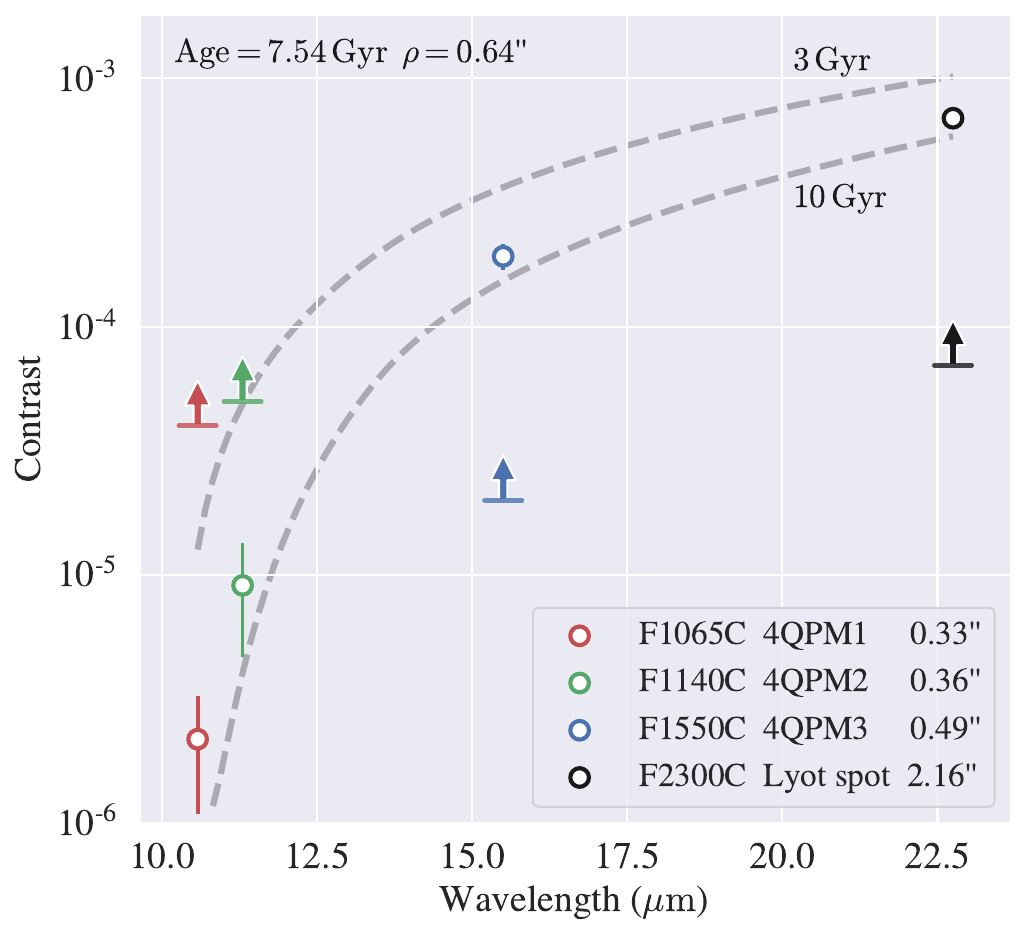}
    \caption{Contrast ratio with respect to MIRI coronagraphy filters. The legend shows the filter, coronagraph and the IWA of MIRI, while the colored circles with error bar represent the estimated contrast ratios assuming an age of $\rm 7.54\pm0.87\,Gyr$. The ratios for F1065C, F1140C, F1550C and F2300C are $2.2(1.1)\times10^{-6}$, $9.1(4.4)\times10^{-6}$, $1.9(0.2)\times10^{-4}$, and $6.9(0.5)\times10^{-4}$, respectively (the numbers in parentheses denote the associated errors). The upward-pointing arrows denote the measured contrast limits by \citet{Boccaletti2022}. The grey dashed lines denote the conservative estimates of contrast ratios assuming a system age of $\rm 3\,Gyr$ and $\rm 10\,Gyr$, respectively. F1550C has contrast ratio and IWA larger than the limit, even if the age of the system ranges from 3 to 10\,Gyr.}
    \label{fig:contrast}
\end{figure}

In this paper, we report the discovery of a long-period and eccentric super-Jupiter HD\,222237\,b located $11.445\pm0.002$\,pc away from our solar system, based on combined analyses of RV, Hipparcos and multiple epochs of Gaia astrometry. 
The planet has $P={40.8}_{-4.5}^{+5.8}$\,yr, $e={0.56}_{-0.03}^{+0.03}$, $i={49.9}_{-2.8}^{+3.4}\degr$, and $m_{\rm p}={5.19}_{-0.58}^{+0.58}\,M_{\rm Jup}$.
Compared with \texttt{orvara} that only utilizes proper-motion anomalies data, our methodology with multiple Gaia data releases being incorporated can avoid the ambiguity of inclination. 
Consequently, we highlight the advantage of our approach for characterizing the orbital properties of cold Jupiters. 

There are some possible caveats about our method. We note that there may be unknown biases associated with using both GDR2 and GDR3. But their solutions are relatively independent apart from the common data they share. On the other hand, correlated data can be used to detect signals if the correlation is well modelled such as how we detect planets in RV data polluted by time-correlated noise. All the noise in GDR2 and GDR3 would not significantly influence our results if they are not significantly time-correlated. 
Unlike our previous work of \citet{Feng2023MNRAS} that uses all of the GOST predictions to model Gaia abscissae, we have taken into account the publicly available gaps (or satellite dead times\footnote{\url{https://www.cosmos.esa.int/web/gaia/dr3-data-gaps}}) in this work. To further validate the impact of these gaps, we conducted a test to compare the mass of HD\,222237\,b derived from solutions with and without correcting for the gaps. The difference in planet mass is found to be relatively small ($0.28\,M_{\rm Jup}$), which is within the $1\,\sigma$ uncertainty reported in this work. 
Regarding the assumption that all Gaia abscissae have the same uncertainty, we note that this is reasonable as long as the uncertainties are not significantly time-dependent. While this assumption may affect the precision of the orbital solution, it only becomes significant when the required precision is well below $\sim1\%$ (see Section 6.1.1 of \citealt{GMBrandt2021AJ}).

Fig.~\ref{fig:mass_ecc_au} shows the location of the planet in eccentricity$-$semi-major axis and mass-angular separation spaces. 
As can be seen in the left panel, HD\,222237\,b has a relatively eccentric orbit in comparison to any exoplanets discovered at large separations ($\rm >10\,au$).
Additionally, the large angular separation (right panel) places it among the small number of cold giant planets that are amenable to further direct imaging characterizations. For the next few years, the planet will continue to move away from its host (Fig.~\ref{fig:orbit}), presenting an excellent opportunity to perform such imaging.
Using the \texttt{ATMO 2020} cooling tracks\footnote{\url{https://perso.ens-lyon.fr/isabelle.baraffe/ATMO2020/}} with the assumption of chemical equilibrium 
for the planetary atmosphere \citep{Phillips2020}, we estimate an effective temperature of $\rm 217\pm6\,K$ ($\lambda_{\rm max}\sim 13.3\,\mu \rm m$) for HD\,222237\,b by adopting an age of $\rm 7.54\pm0.87\,Gyr$ \citep{Lovis2011} derived through the activity-rotation-age calibration \citep{Mamajek2008}. 
We also estimate an upper limit of equilibrium temperature of 61\,K using
\begin{equation}
    T_{\rm eq}=T_{\rm eff\star}(1-A_B)^{1/4}\sqrt{\frac{R_\star}{2a}},
\end{equation}
where $A_B$ is the Bond albedo ($A_B=0$), $a$ is the semi-major axis of the planet, and $T_{\rm eff\star}$ and $R_\star$ are stellar effective temperature and radius. Then by assuming a blackbody radiation for the host star HD\,222237\,, we can calculate its apparent (Vega) magnitude in different bandpass according to the filter response\footnote{Download from the SVO Filter Profile Service \citep{Rodrigo2020}: \url{http://svo2.cab.inta-csic.es/theory/fps3/}}, the Vega spectrum and the distance. While based on the planetary mass and the system age, the absolute magnitude of the planet can be directly obtained through the interpolation\footnote{Using the python \texttt{scipy.interpolate.interp1d} library} of the cooling tracks (\texttt{ATMO 2020} models provide pre-calculated absolute magnitudes in a number of common photometric filters). These absolute magnitudes will be converted to apparent magnitudes assuming the planet has the same distance as the star. Ultimately, the contrast ratios of the planet to its host in different filters can be inferred.

We found the contrast ratios in $J$, $H$, $K$ bands are as low as $\sim10^{-10}$, significantly lower than the typical contrast limit ($\sim10^{-6}$) of the current ground-based coronagraphs, such as SCExAO/CHARIS installed on the Subaru telescope \citep{Jovanovic2015, Currie2020}. This means the planet is undetectable in near-infrared band by those facilities, but it may be detectable using the Mid-Infrared Instrument (MIRI) \citep{Rieke2015} mounted on JWST. 
Fig.~\ref{fig:contrast} shows the derived contrast ratios of the system in different JWST/MIRI coronagraph filters. 
It is significant that the inner working angles (IWA) of three four-quadrant phase masks (4QPM; \citealt{Rouan2000}) coronagraphs are smaller than the planet-star separation at epoch 2025.0. 
Furthermore, when comparing the actual performance (\citealt{Boccaletti2022}, see their Fig. 5) of the MIRI coronagraphs, we found the F1550C filter with reference star subtraction and long integrations seems appropriate for imaging the planet, even if we adopt a more broader assumption of the system age. 

In addition, we explore the contribution of planetary reflected light on the contrast. The flux ratio of the planet to the host star is expressed as (e.g., \citealt{Kane2010ApJ}, Equation 7) 
\begin{equation}
    \frac{f_{\rm p}(\phi, \lambda)}{f_\star(\lambda)}=A_{\rm g}(\lambda)g(\phi, \lambda)\frac{R_{\rm p}^2}{r^2},
\end{equation}
where $\phi$ is the phase angle, $A_{\rm g}(\lambda)$ is the geometric albedo ranging from 0 to 1, $g(\phi, \lambda)$ is the phase function ranging from 0 to 1, $r$ is the distance to the star, and $R_{\rm p}$ is the planetary radius that can be obtained by interpolating the cooling tracks. Assuming $g=1$ and $A_{\rm g}=0.5$, the magnitude of the contrast ratio is estimated to be $\sim10^{-9}$, larger than those derived from cooling model in near-infrared band, but it remains undetectable.
Nevertheless, this suggests that the thermal emission of the planet in the near-infrared band might be dominated by the reflection of starlight instead of self-luminosity.

\citet{Bowler2020} suggested that, based on their population-level eccentricity analysis examining directly-imaged substellar companions, companions with $M_{\rm p}<15\,M_{\rm Jup}$ tend to have relatively lower orbital eccentricity, while brown dwarfs (BDs) exhibit higher eccentricity. The authors interpreted this as evidence for imaged planets formed through core accretion, and for BDs formed through molecular cloud fragmentation. 
We note, however, that HD\,222237\,b has an eccentricity of ${0.56}_{-0.03}^{+0.03}$, implying the possibility of experiencing some kinds of severe orbital evolution, such as planet-planet scattering (e.g., \citealt{Ford2008}) or perturbations from third-body fly-by (e.g., \citealt{Naoz2016}). Furthermore, the metal-poor condition ($-0.32$\,dex) along with the wide separation appear to contradict the predictions of the core accretion paradigm (e.g., \citealt{Ida2004,Mordasini2012}), although we can't rule out that the core of the planet initially formed on small separation and then underwent outward scattering and runaway accretion \citep{Marleau2019}. 
On the other hand, disc instability is thought to be metal-independent and occur far away from the central star ($>10\,{\rm au}$), where it allows for more efficient cooling and collapse, resulting in the formation of massive companions (e.g., \citealt{Rice2022, Meru2010}). Some works pointed out that giant planets might not prefer orbiting metal-rich hosts above a limit of $\sim4\,M_{\rm Jup}$, i.e., more massive planets might show similar formation channel with BDs (e.g., \citealt{Santos2017, Schlaufman2018, Maldonado2019}). The population synthesis model \citep{Forgan2018} predicts that it is possible for some massive companions to undergo inward migration \citep{Baruteau2011} and tidal disruption \citep{Nayakshin2017} to decrease their mass on a much closer-in and eccentric orbit \citep{Rice2022}. In conclusion, it is likely that disc instability is responsible for the formation of HD\,222237\,b, but we can't exclude a formation by core accretion at small separation.
This system would benefit from high contrast imaging studies to disentangle the truth from the ambiguities of its formation and dynamics.

\section*{Acknowledgements}
We thank the anonymous referees for their insightful comments and valuable suggestions that greatly improved our paper.
This research is supported by Shanghai Jiao Tong University 2030 Initiative. 
This research is also supported by the National Natural Science Foundation of China (NSFC) under Grant No. 12473066, and the Chinese Academy of Sciences President’s International Fellowship Initiative grant No. 2020VMA0033. M.R. acknowledges support from Heising-Simons Foundation Grant \#2023-4478.
The authors acknowledge the years of technical support from LCO staff in the successful operation of PFS, enabling the collection of the data presented in this paper. We also thank Samuel W. Yee for his effort on the observations and Pablo Pe\~{n}a for his valuable cross-check of orbital period.
This work is based in part on data acquired at the Anglo-Australian Telescope. We acknowledge the traditional custodians of the land on which the AAT stands, the Gamilaraay people, and pay our respects to elders past and present. 
The computations in this paper were run on the $\pi$ 2.0 (or the Siyuan-1) cluster supported by the Center for High Performance Computing at Shanghai Jiao Tong University.

This research has made use of the SIMBAD database, operated at CDS, Strasbourg, France \citep{Wenger2000}.
This work presents results from the European Space Agency (ESA) space mission Gaia. Gaia data are being processed by the Gaia Data Processing and Analysis Consortium (DPAC). Funding for the DPAC is provided by national institutions, in particular the institutions participating in the Gaia MultiLateral Agreement (MLA). The Gaia mission website is \href{https://www.cosmos.esa.int/gaia}{https://www.cosmos.esa.int/gaia}. The Gaia archive website is \href{https://archives.esac.esa.int/gaia}{https://archives.esac.esa.int/gaia}. This paper is partly based on observations collected at the European Organisation for Astronomical Research in the Southern Hemisphere under ESO programmes: 192.C-0852, 072.C-0488, 183.C-0972.

\section*{Data Availability}
The PFS and AAT RV data are available in the appendix while other data are publicly available.



\bibliographystyle{mnras}
\bibliography{example} 




\appendix

\section{RVs for HD 222237}
\label{APPE_A}
\begin{table*}
\centering
\caption{PFS RVs for HD\,222237}\label{Tab:RV_222237}
\begin{tabular*}{\textwidth}{@{}@{\extracolsep{\fill}}cccccccc@{}}
\hline \hline
BJD & RV [$\rm m\,s^{-1}$] & Error [$\rm m\,s^{-1}$] & $S$-index & BJD & RV [$\rm m\,s^{-1}$] & Error [$\rm m\,s^{-1}$] & $S$-index\\
\hline
2455785.74573&-30.19&0.91&0.227&2458717.64622&47.90&0.79&0.212\\
2455787.83202&-30.82&0.96&0.186&2458738.67411&48.70&0.89&0.203\\
2455787.83472&-31.86&0.95&0.187&2458738.67807&47.74&0.92&0.201\\
2455790.76822&-32.10&1.04&0.257&2458744.68167&49.47&1.10&0.267\\
2455790.77201&-29.42&1.02&0.255&2458744.68998&46.72&1.15&0.251\\
2455793.79756&-29.46&0.93&0.246&2459153.66133&29.66&0.79&0.222\\
2455796.77370&-28.75&1.03&0.246&2459153.66421&30.49&0.81&0.232\\
2455796.77832&-29.11&1.19&0.263&2459153.66720&30.10&0.81&0.224\\
2456139.75501&-23.85&1.06&0.248&2459363.89612&24.41&0.94&0.211\\
2456175.76174&-26.57&1.04&0.262&2459363.89946&24.76&0.92&0.223\\
2456176.72566&-28.48&1.03&0.234&2459447.69813&14.24&0.83&0.214\\
2456501.82731&-14.66&0.96&0.239&2459447.70012&15.71&0.87&0.205\\
2456550.66118&-18.60&0.92&0.249&2459447.70211&15.95&0.82&0.205\\
2456552.67848&-17.30&1.10&0.240&2459447.70413&16.88&0.81&0.212\\
2456604.60810&-14.66&0.79&0.241&2459507.67315&12.93&0.89&0.203\\
2456607.58228&-10.50&1.02&0.237&2459507.67574&11.37&0.79&0.211\\
2456818.94017&-8.20&0.99&0.231&2459507.67833&11.76&0.87&0.208\\
2456867.83024&-11.15&1.01&0.214&2459748.92057&3.25&0.97&0.227\\
2456871.77590&-11.73&0.87&0.220&2459748.92316&3.14&0.91&0.227\\
2457260.79250&0.00&0.88&0.246&2459748.92580&4.47&0.88&0.227\\
2457319.66988&-1.44&0.90&0.235&2459829.72691&-0.71&0.97&0.223\\
2457326.61668&-2.51&0.88&0.233&2459829.72941&-1.13&0.94&0.221\\
2457536.93357&6.80&1.09&0.244&2459829.73202&-1.42&0.95&0.225\\
2457555.92679&7.15&0.90&0.228&2459861.65634&0.18&0.86&0.214\\
2457616.77039&10.82&0.87&0.212&2459861.65892&-1.42&0.81&0.214\\
2457619.78799&8.38&1.11&0.342&2459861.66147&-0.91&0.94&0.227\\
2457737.54332&12.65&0.88&0.218&2459890.54516&-4.26&0.92&0.240\\
2458271.85419&44.44&0.90&0.210&2459890.55021&-3.55&0.94&0.236\\
2458293.88374&39.93&0.81&0.204&2460124.93080&-7.41&0.88&0.223\\
2458675.82705&47.65&0.89&0.220&2460124.93330&-9.37&0.92&0.218\\
2458675.83079&47.53&0.98&0.226&2460124.93589&-7.52&0.80&0.215\\
2458675.83326&47.09&0.92&0.227&&&&\\
\hline
\end{tabular*}
\end{table*}

\begin{table*}
\centering
\caption{AAT RVs for HD\,222237}\label{Tab:RV_222237_AAT}
\begin{tabular*}{\textwidth}{@{}@{\extracolsep{\fill}}cccccc@{}}
\hline \hline
BJD & RV [$\rm m\,s^{-1}$] & Error [$\rm m\,s^{-1}$]  & BJD & RV [$\rm m\,s^{-1}$] & Error [$\rm m\,s^{-1}$] \\
\hline
2451034.23140&-4.92&1.93&2453042.91623&0.89&1.94\\
2451385.32779&-14.86&3.34&2453245.21460&-0.50&1.62\\
2451473.08162&-0.28&2.24&2453570.24266&-3.32&1.44\\
2451745.25030&-8.45&2.42&2453947.27765&-0.36&1.26\\
2451830.07472&1.99&4.38&2454041.05363&-2.61&1.49\\
2451920.94390&-2.50&2.50&2454255.22288&-1.10&1.50\\
2452128.21848&-1.44&1.76&2454371.18539&1.63&1.28\\
2452152.02008&-10.86&2.21&2454777.07424&-2.18&1.10\\
2452187.10631&-6.09&1.35&2455106.14845&1.23&1.87\\
2452456.27032&-5.94&1.78&2455846.00705&15.31&1.49\\
2452511.05088&-2.23&1.53&2456140.23712&25.33&1.83\\
2452591.98274&-3.04&1.06&2456465.31881&35.88&1.77\\
2452593.98191&-0.54&1.77&2456556.08965&29.32&1.45\\
2452594.99536&-4.84&1.56&2456935.12973&34.35&1.60\\
2452598.98813&-4.48&1.82&2456939.09431&21.83&1.87\\
2452787.29273&-1.46&2.78&2457236.26158&51.22&2.73\\
2452861.27062&-0.28&2.55&2457346.01176&44.06&2.34\\
2452945.08777&-1.35&1.82&&\\
\hline
\end{tabular*}
\end{table*}

\section{DETAILED METHODOLOGY}
\label{APPE_B}
\subsection{RV Model}
For an elliptical orbit, the distance of a star from the system's barycenter and the star's z-coordinate along the line-of-sightis are respectively 
\begin{equation}
r(t)=\frac{a_{\star}(1-e^2)}{1+e\,\cos\nu(t)}~,~{\rm and}
\end{equation}
\begin{equation}
z(t)=r(t)\sin i\,\sin(\omega+\nu(t))~,
\end{equation}
where $a_{\star}$ is the semi-major axis of the primary star relative to the system's barycenter, $e$ is the eccentricity, $i$ is the inclination, $\omega$ is the argument of periastron of the stellar reflex motion, and $\nu(t)$ is the true anomaly and is related to the eccentric anomaly, $E(t)$, which is given by 
\begin{equation}
    {\rm tan}\frac{\nu(t)}{2}=\sqrt{\frac{1+e}{1-e}}\cdot {\rm tan}\frac{E(t)}{2}~.
\end{equation}
This relation can be derived geometrically.
The mean anomaly $M(t)$ at a specific time is then defined as
\begin{equation}
    M(t)=\frac{2\pi}{P}(t-T_{p})~.
\end{equation} 
According to Kepler’s equation, the relation between $M(t)$ and $E(t)$ is given by
\begin{equation}
    M(t)=E(t)-e\,{\rm sin}\,E(t)~.
\end{equation}
Thus, the variation of stellar RVs due to a companion at epoch $t_j$ is
\begin{equation}
  \hat{v}_{j}=\dot{z}= K[\cos(\omega+\nu(t_j))+e\cos(\omega)]~,
\end{equation}
where $K$ is the semi-amplitude and can be written as
\begin{equation}
K\equiv\frac{2\pi}{P}\frac{a_{\star}{\rm sin}\,i}{\sqrt{1-e^2}}.
\label{eq:K}
\end{equation}
Therefore, the likelihood for the measured RV ($v_{j,k}$) can be calculated by
\begin{equation}
  \mathcal{L}_{\rm RV}=\prod\limits_{j=1}^{N_{\rm RV}}\prod\limits_{k=1}^{N_{\rm inst}}\frac{1}{\sqrt{2\pi(\sigma_{j,k}^2+\sigma_{{\rm jit},k}^2)}}{\rm exp}\left(-\frac{(v_{j,k}-\hat{v}_{j,k}-\gamma_k)^2}{2(\sigma_{j,k}^2+\sigma_{{\rm jit},k}^2)}    \right)~,
\end{equation}
where $N_{\rm RV}$ and $N_{\rm inst}$ are respectively the number of RV measurements and instruments, and $\gamma_k$ and $\sigma_{{\rm jit},k}$ are respectively the RV offset and the so-called ``RV Jitter'' for different instruments.

\subsection{Gaia Astrometric Model}
In rectangular coordinates, the Thiele-Innes coefficients $A$, $B$, $F$, $G$ are defined as
\begin{equation}
    A={\rm cos}\,\omega\,{\rm cos}\,\Omega -{\rm sin}\,\omega\,{\rm sin}\,\Omega\,{\rm cos}\,i,
\end{equation}
\begin{equation}
    B={\rm cos}\,\omega\,{\rm sin}\,\Omega +{\rm sin}\,\omega\,{\rm cos}\,\Omega\,{\rm cos}\,i,
\end{equation}
\begin{equation}
    F=-{\rm sin}\,\omega\,{\rm cos}\,\Omega -{\rm cos}\,\omega\,{\rm sin}\,\Omega\,{\rm cos}\,i,
\end{equation}
\begin{equation}
    G=-{\rm sin}\,\omega\,{\rm sin}\,\Omega +{\rm cos}\,\omega\,{\rm cos}\,\Omega\,{\rm cos}\,i,
\end{equation}
where $\Omega$ is the
longitude of the ascending node.
Besides, the elliptical rectangular coordinates $X$ and $Y$ are functions of the eccentric anomaly $E(t)$ and the eccentricity $e$, which are given by
\begin{equation}
    X={\rm cos}\,E(t)-e
\end{equation}
\begin{equation}
    Y=\sqrt{1-e^2}\cdot {\rm sin}\,E(t).
\end{equation}
Therefore, the projected offsets of stellar reflex motion relative to the system’s barycenter are then given by
\begin{equation}
    \Delta\alpha^{r}_\ast=a_{\star}\varpi(BX+GY),
\end{equation}
\begin{equation}
     \Delta\delta^{r}=a_{\star}\varpi(AX+FY),
\end{equation}
where $\Delta\delta^{r}$ and $\Delta\alpha^{r}_\ast=\Delta\alpha^r\,{\rm cos}\,\delta^r$ are the offsets in declination and right ascension, respectively, and $\varpi$ is the parallax in units of mas. It is noted that we assume the companion is fainter than its host, and therefore its luminosity contributed to the system's photocenter is negligible (photocenter is equal to barycenter).
Next, we model the astrometry of TSB at the GDR3 epoch ($t_{\rm DR3}={\rm J}2016.0$) as follows:
\begin{equation}
    \alpha^{b}_{\rm DR3}=\alpha_{\rm DR3}-\frac{\Delta\alpha_\ast}{{\rm cos}\,\delta_{\rm DR3}}
\end{equation}
\begin{equation}
    \delta^{b}_{\rm DR3}=\delta_{\rm DR3}-\Delta\delta
\end{equation}
\begin{equation}
    \varpi^{b}_{\rm DR3}=\varpi_{\rm DR3}-\Delta\varpi
\end{equation}
\begin{equation}
    \mu^{b}_{\alpha{\rm DR3}}=\mu_{\alpha{\rm DR3}}-\Delta\mu_{\alpha}
\end{equation}
\begin{equation}
    \mu^{b}_{\delta{\rm DR3}}=\mu_{\delta{\rm DR3}}-\Delta\mu_{\delta},
\end{equation}
where $\alpha$, $\delta$, $\mu_{\alpha}$, $\mu_{\delta}$ are right ascension, declination and corresponding proper motions, and the subscript $_{\rm DR3}$ and the superscript $^b$ represent quantities of GDR3 and TSB astrometry, respectively.
Above five quantities with $\Delta$ are barycenter offsets relative to GDR3 astrometry, and will be set as free parameters.
The TSB astrometry at reference epoch $t_k$ ($k=1,2$ represent GDR2, GDR3) can be modeled through linear propagation in the Cartesian coordinate as follows \citep{Lindegren2012,Feng2019ApJS}:
\begin{equation}
\left[\begin{array}{c}
     x  \\
     y  \\
     z
\end{array}\right]=d\,
\left[\begin{array}{c}
     \cos\alpha^b\,\cos\delta^b  \\
     \sin\alpha^b\,\cos\delta^b  \\
     \sin\delta^b
\end{array}\right],~
\left[\begin{array}{c}
     v_x  \\
     v_y  \\
     v_z
\end{array}\right]=
\left[\begin{array}{ccc}
     \cos\delta^b\,\cos\alpha^b&-\sin\delta^b\,\cos\alpha^b&-\sin\alpha^b  \\
     \cos\delta^b\,\sin\alpha^b&-\sin\delta^b\,\sin\alpha^b&\cos\alpha^b  \\
     \sin\delta^b&\cos\delta^b&0
\end{array}\right]
\left[\begin{array}{c}
     v_r\\
     v_\delta\\
     v_\alpha
\end{array}\right],~
\left[\begin{array}{c}
     x_k\\
     y_k\\
     z_k
\end{array}\right]=
\left[\begin{array}{c}
     x\\
     y\\
     z
\end{array}\right]+\Delta t_k
\left[\begin{array}{c}
     v_r\\
     v_\delta\\
     v_\alpha
\end{array}\right]~,
\end{equation}
and 
\begin{equation}
\left[\begin{array}{c}
     v^b_{rk}\\
     \mu^b_{\alpha k} d_k\\
     \mu^b_{\delta k} d_k
\end{array}\right]=
\left[\begin{array}{ccc}
     \cos\delta_k^b&0&\sin\delta_k^b\\
     0&1&0\\
     -\sin\delta_k^b&0&\cos\delta^b
\end{array}\right]
\left[\begin{array}{ccc}
     \cos\alpha_k^b&\sin\alpha_k^b&0\\
     -\sin\alpha_k^b&\cos\alpha_k^b&0\\
     0&0&1
\end{array}\right]
\left[\begin{array}{c}
     v_x\\
     v_y\\
     v_z
\end{array}\right],
\end{equation}
where $d=1/\varpi^b$, $v_\delta=\mu_\delta^b d$, $v_\alpha=\mu_\alpha^b d$, $v_r\approx \rm RV_{DR3}$, $d_k=\sqrt{x_k^2+y_k^2+z_k^2}=1/\varpi_k^b$, and $\Delta t_k$ represents the difference between $t_k$ and $t_{\rm DR3}$. 
Once we obtain the Cartesian state vector ($x_k,y_k,z_k,v_x,v_y,v_z$) at $t_k$, we can transform them back to the equatorial state vector ($\alpha_k^b, \delta_k^b, \mu^b_{\alpha k}, \mu^b_{\delta k}, \varpi_k^b, v^b_{rk}$). Since the GDR3 RVs are not precise enough to constrain the reflex motion, we only use them to propagate astrometry of TSB. This propagation in Cartesian coordinate system instead of spherical coordinate system can effectively avoid the problem of perspective acceleration.

By combining the linear motion of TSB and the target reflex motion in the equatorial coordinate system, we can simulate the Gaia and Hipparcos AL abscissae directly. To obtain GDR3 abscissae, we first simulate the position of the target at GOST epoch $t_j$ relative to $t_{\rm DR3}$ using 
\begin{equation}
    \Delta\alpha_{\ast{j}}=\Delta\alpha^{b}_{\ast{\rm DR3}}+\mu^{b}_{\alpha {\rm DR3}}(t_j-t_{\rm DR3})+\Delta\alpha^{r}_{\ast j}
\end{equation}
\begin{equation}
    \Delta\delta_{j}=\Delta\delta^{b}_{\rm DR3}+\mu^{b}_{\delta {\rm DR3}}(t_j-t_{\rm DR3})+\Delta\delta^{r}_{j}
\end{equation}
where $\Delta\alpha^{b}_{\ast{\rm DR3}}=(\alpha^b_{\rm DR3}-\alpha_{\rm DR3})\,\cos\delta^b_{\rm DR3}$, and $\Delta\delta^{b}_{\rm DR3}=\delta^b_{\rm DR3}-\delta_{\rm DR3}$. Since the reflex motion induced by substellar companions is not as significant as linear barycentric motion, we approximate the parallax at $t_j$ as $\varpi_j\approx\varpi^b_{\rm DR3}$. Then we project the above target position onto the 1D AL direction by considering the parallactic perturbation of Gaia satellite's heliocentric motion, using 
\begin{equation}
    \eta_j = \Delta\alpha_{\ast{j}}\,{\rm sin}\,\psi_j+\Delta\delta_{j}\,{\rm cos}\,\psi_j+\varpi^{b}_{\rm DR3}f^{\rm AL}_j,~
    \label{equ:gaia_abs}
\end{equation}
where $\eta_j$ is AL abscissa, $\psi_j$ is the scan angle of Gaia satellite, and $f^{\rm AL}_j$ is the parallax factor from GOST. Finally, we model the simulated abscissae with a five-parameter model as follows:
\begin{equation}
    \hat{\eta_j} = \Delta\alpha^{l}_{\ast{\rm DR3}}\,{\rm sin}\,\psi_j+\Delta\delta^{l}_{\rm DR3}\,{\rm cos}\,\psi_j+\hat{\varpi}_{\rm DR3}f^{\rm AL}_j,
\end{equation}
\begin{equation}
\Delta\alpha^{l}_{\ast{\rm DR3}}=(\hat{\alpha}_{\rm DR3}-\alpha_{\rm DR3})\,{\rm cos}\,\hat{\delta}_{\rm DR3}+\hat{\mu}_{\alpha{\rm DR3}}(t_j-t_{\rm DR3}),
\end{equation}
\begin{equation}
\Delta\delta^{l}_{\rm DR3}=(\hat{\delta}_{\rm DR3}-\delta_{\rm DR3})+\hat{\mu}_{\delta{\rm DR3}}(t_j-t_{\rm DR3}).
\end{equation}
Above modelling can give a set of model parameters ($\hat{\alpha}_{\rm DR3}, \hat{\delta}_{\rm DR3}, \hat{\mu}_{\alpha{\rm DR3}},   \hat{\delta}_{\rm DR3}, \hat{\varpi}_{\rm DR3}$) at $t_{\rm DR3}$. Likewise, modeling GDR2 astrometry can be done easily by changing the subscript $_{\rm DR3}$ to $_{\rm DR2}$, but keeping the reference position fixed in GDR3. Given that the Gaia IAD is not available, we assume each individual observation has the same uncertainty and thus will be assigned equal weighting when fitting for the five-parameter model. 
Besides, we take into account the published astrometric gaps (e.g., dead times and rejected observations) when modelling abscissae. 
To avoid numerical errors, we define the catalog astrometry at $t_k$ relative to $t_{\rm DR3}$ as follow:
\begin{equation}
\Delta\vec{\iota}_k\equiv(\Delta\alpha_{\ast k},\,\Delta\delta_k,\,\Delta\varpi_k,\,\Delta\mu_{\alpha k},\,\Delta\mu_{\delta k})=((\alpha_k-\alpha_{\rm DR3})\,{\rm cos\delta_k},\,\delta_k-\delta_{\rm DR3},\,\varpi_k-\varpi_{\rm DR3},\,\mu_{\alpha k}-\mu_{\alpha \rm DR3},\,\mu_{\delta k}-\mu_{\delta \rm DR3}).
\end{equation}
Likewise, the fitted astrometry at $t_k$ is $\Delta\hat{\vec{\iota}}_k$. The likelihood for GDR2 and GDR3 can be written as
\begin{equation}
  \mathcal{L}_{\rm gaia}=\prod\limits_{k=1}^{N_{\rm DR}}\frac{1}{\sqrt{(2\pi)^5|\Sigma_k(S^2)|}}{\rm exp}\left(-\frac{1}{2}(\Delta\hat{\vec{\iota}}_k-\Delta\vec{\iota}_k)^{T}[\Sigma_k(S^2)]^{-1}(\Delta\hat{\vec{\iota}}_k-\Delta\vec{\iota}_k)    \right)~,
\end{equation}
where $N_{\rm DR}$ represents the number of Gaia data releases ($N_{\rm DR}=2$ if we use both GDR2 and GDR3), $\Sigma_k$ is the catalog covariance for the five parameters, and $S$ is the error inflation factor for Gaia astrometry. 
Given that the covariance given by Gaia catalog is probably underestimated, we can use the error inflation $S$ and jitter $J$ to construct a new covariance as $\Sigma_{mn}=\rho_{mn}\sqrt{S^2\sigma_n^2+J^2}\sqrt{S^2\sigma_k^2+J^2}$, where $\rho$ is the correlation matrix.
As indicated by \citet{Feng2024ApJS} (see their Table 1) who employs orbital solutions from the GDR3 non-single-star catalog (NSS; \citealt{Halbwachs2023A&A, Holl2023A&A}) to estimate the error inflation within the astrometric catalogs, no significant jitter and error inflation are found existing in both GDR2 and GDR3, and this finding remains robust across different choices of calibration sources. We thus use a strong Gaussian distribution as the prior to constrain the error inflation, as well as setting jitter to zero. 

\subsection{Hipparcos Astrometric Model}
Similar to Gaia astrometric model, we first propagate the TSB astrometry at $t_{\rm DR3}$ to Hipparcos reference epoch $t_{\rm HIP}$. Then we simulate the position of target at Hipparcos epoch using
\begin{equation}
    \Delta\alpha_{\ast{j}}=\Delta\alpha^b_{\ast{\rm HIP}}+\Delta\mu^b_{\alpha {\rm HIP}}(t_j-t_{\rm HIP})+\Delta\alpha^{r}_{\ast j},
\end{equation}
\begin{equation}
    \Delta\delta_{j}=\Delta\delta^b_{\rm HIP}+\Delta\mu^b_{\delta {\rm HIP}}(t_j-t_{\rm HIP})+\Delta\delta^{r}_{j},
\end{equation}
where $\Delta\alpha^{b}_{\ast{\rm HIP}}=(\alpha^b_{\rm HIP}-\alpha_{\rm HIP})\,\cos(\Delta\delta^b_{\rm HIP}/2)$, $\Delta\delta^{b}_{\rm HIP}=\delta^b_{\rm HIP}-\delta_{\rm HIP}$, $\Delta\mu^b_{\alpha {\rm HIP}}=\mu^b_{\alpha {\rm HIP}}-\mu_{\alpha {\rm HIP}}$, and $\Delta\mu^b_{\delta {\rm HIP}}=\mu^b_{\delta {\rm HIP}}-\mu_{\delta {\rm HIP}}$. Therefore, the abscissae of Hipparcos is given by
\begin{equation}
    \hat{\xi_j} = \Delta\alpha_{\ast{j}}\,{\rm cos}\,\psi_j+\Delta\delta_{j}\,{\rm sin}\,\psi_j+\Delta\varpi^{b}_{\rm HIP}f^{\rm AL}_j,~ 
\end{equation}
where $\Delta\varpi^{b}_{\rm HIP}=\varpi^{b}_{\rm HIP}-\varpi_{\rm HIP}$. Above three formulae are slightly different with that of Gaia. We additionally correct the difference between Hippacos astrometry and the astrometry propagated from the GDR3 epoch to the Hipparcos epoch.
Besides, the scan angle in the new Hipparcos IAD is complementary with Gaia scan angle. Finally, we can calculate the likelihood for Hipparcos IAD using
\begin{equation}
  \mathcal{L}_{\rm hip}=\prod\limits_{j=1}^{N_{\rm IAD}}\frac{1}{\sqrt{2\pi(\sigma_{j}^2+J_{\rm hip}^2)}}{\rm exp}\left(-\frac{(\hat{\xi}_j-\xi_j)^2}{2(\sigma_{j}^2+J_{\rm hip}^2)}    \right)~,
\end{equation}
where $N_{\rm IAD}$ is the total number of Hipparcos IAD, $\sigma_j$ is the individual measurement uncertainty, and $J_{\rm hip}$ is the jitter term.

\section{ADDITIONAL FIGURES}
\label{APPE_C}

\begin{figure*}
    \centering
	\includegraphics[width=\textwidth]{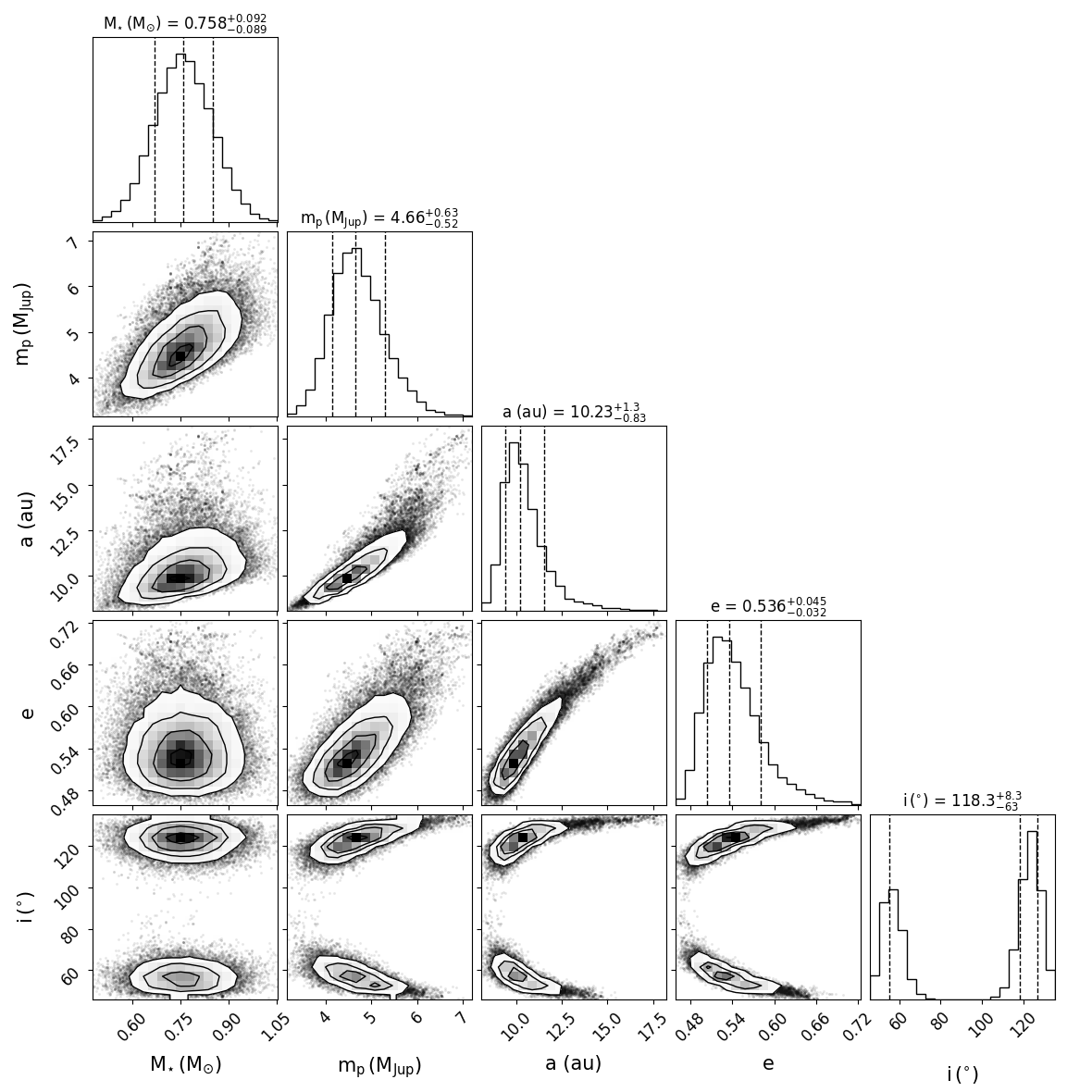}
    \caption{Posterior distributions for selected orbital parameters by \texttt{orvara} (RV+HGCA). The median and the corresponding the $1\sigma$ confidence intervals are denoted by vertical dashed lines. }
    \label{fig:corner_orvara}
\end{figure*}


\begin{figure*}
    \centering
	\includegraphics[width=\textwidth]{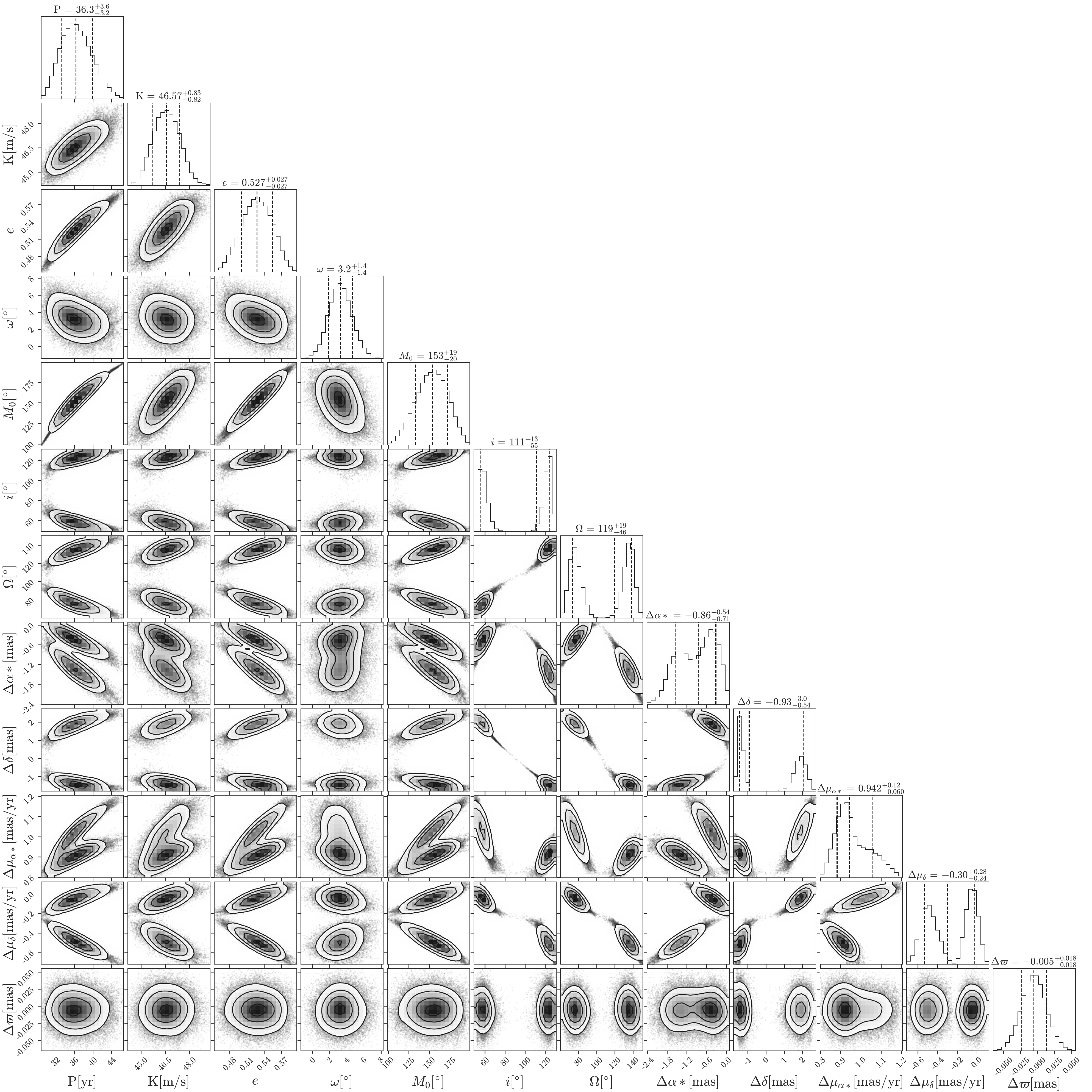}
    \caption{Posterior distributions for selected orbital parameters by RV+HG3. The bimodal distributions of $i$, $\Omega$, $\Delta \alpha*$, $\Delta \delta$, $\Delta \mu_{\alpha*}$ and $\Delta \mu_\delta$ can be recognized. }
    \label{fig:corner_HG3}
\end{figure*}

\begin{figure*}
    \centering
	\includegraphics[width=\textwidth]{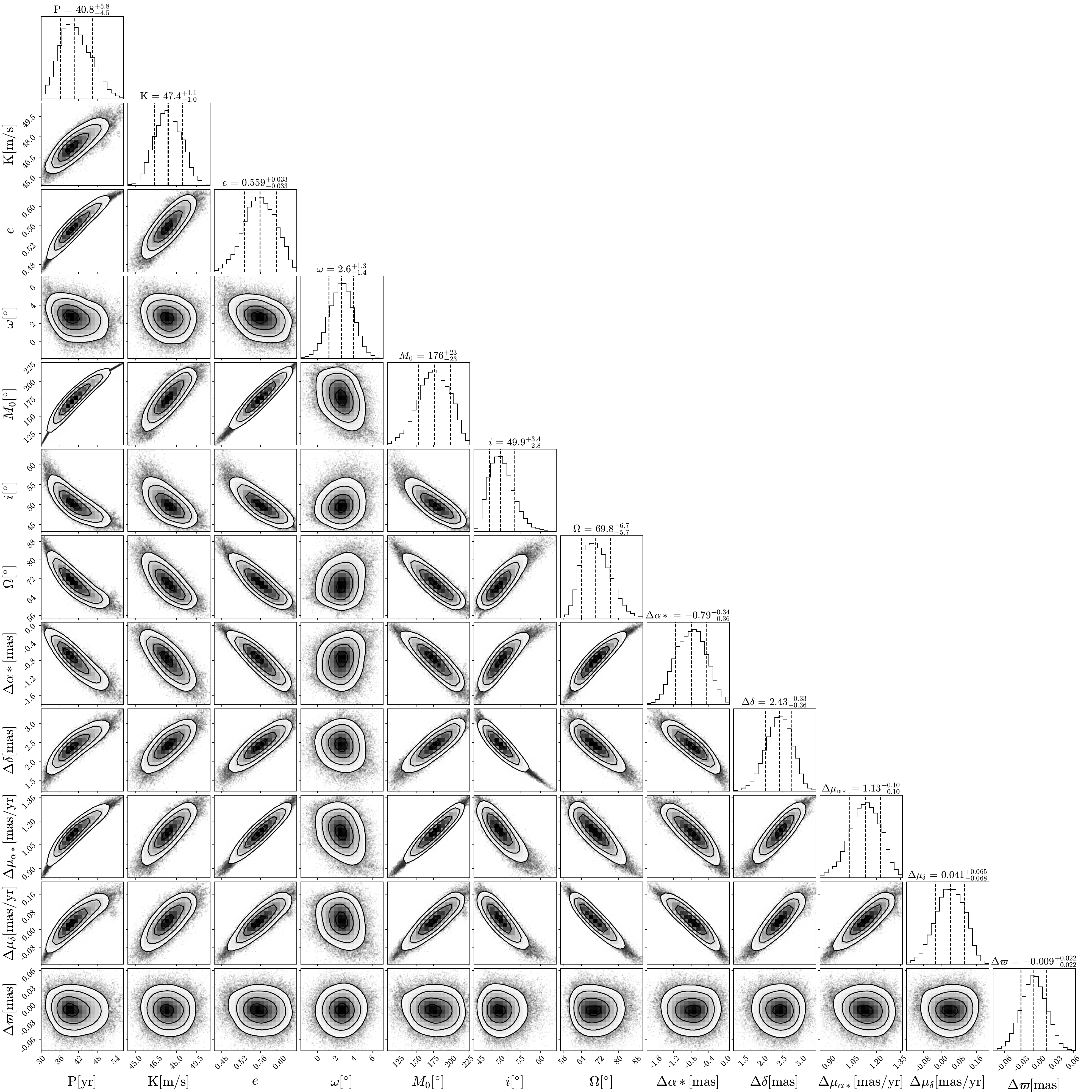}
    \caption{Posterior distributions for selected orbital parameters by RV+HG23. With the inclusion of GDR2, the ambiguities for $i$, $\Omega$, $\Delta \alpha*$, $\Delta \delta$, $\Delta \mu_{\alpha*}$ and $\Delta \mu_\delta$ are well resolved. }
    \label{fig:corner_HG23}
\end{figure*}

\begin{figure*}
\centering
\begin{subfigure}[b]{0.45\textwidth}
  \includegraphics[width=\textwidth]{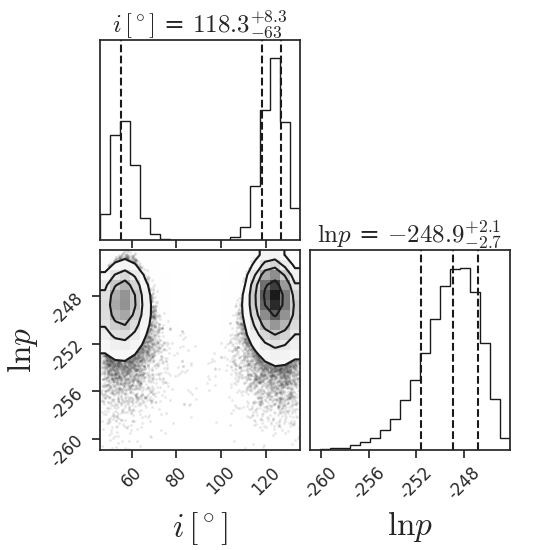}
  \caption{\texttt{orvara (RV+HGCA) solution}}
\end{subfigure}
\hfill
\begin{subfigure}[b]{0.45\textwidth}
  \includegraphics[width=\textwidth]{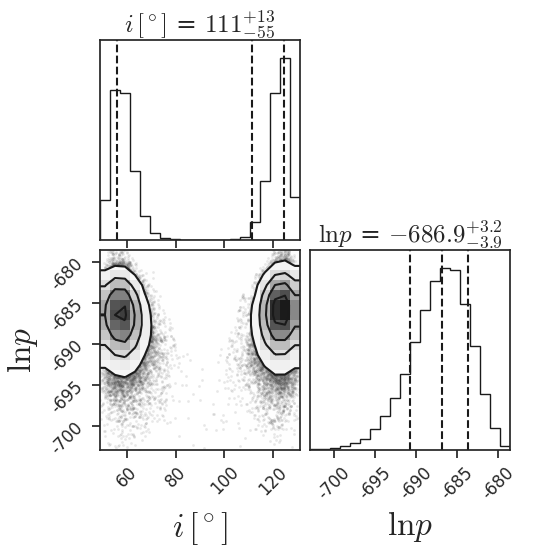}
  \caption{RV+HG3 solution}
\end{subfigure}
\vspace{0.1cm}
\begin{subfigure}[b]{0.45\textwidth}
  \includegraphics[width=\textwidth]{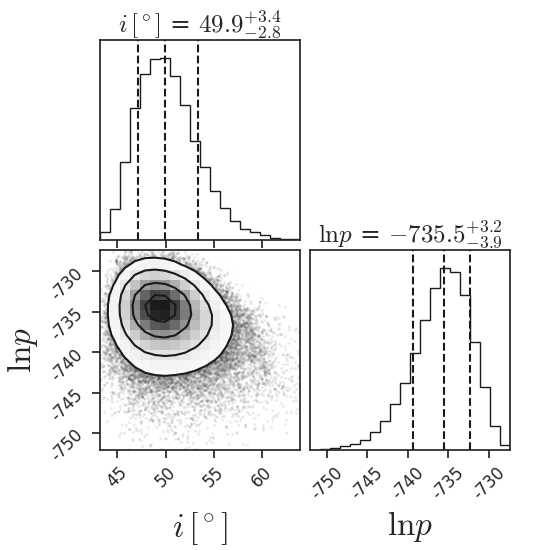}
  \caption{RV+HG23 solution}
\end{subfigure}
\hfill
\begin{subfigure}[b]{0.45\textwidth}
  \includegraphics[width=\textwidth]{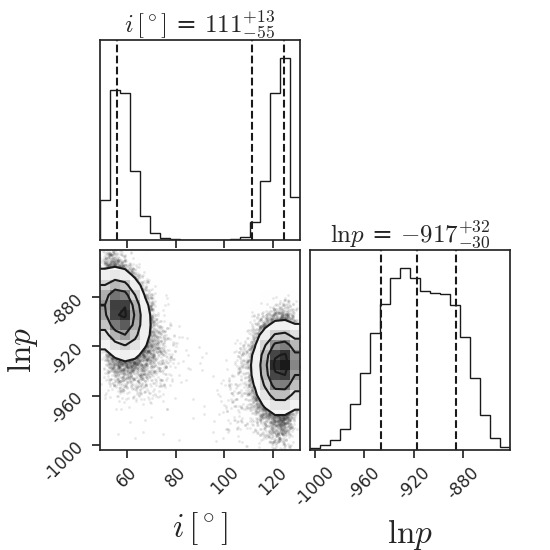}
  \caption{Injecting RV+HG3 solution into RV+HG23 model}
\end{subfigure}
\caption{Comparing the inclination posterior distribution of different solutions. The inclination obtained from \texttt{orvara} (a) and RV+HG3 (b) exhibits bimodal distribution, and the two modes have equivalent posterior probability, while RV+HG23 (c) has the ability to break the inclination degeneracy. When RV+HG3 solution is injected into RV+HG23 model (d), we found that the higher inclination (corresponding to the retrograde orbital solution) can be rejected, suggesting the precision of Gaia, along with the baseline between GDR2 and GDR3 are sufficient to obtain an unambiguous orbital orientation of HD\,222237\,b.} 
\label{fig:fourpics}
\end{figure*}

\bsp	
\label{lastpage}
\end{document}